\documentclass[usenatbib]{basi}
\usepackage[T1]{fontenc}
\usepackage[british]{babel}
\usepackage[varg]{txfonts}
\usepackage{rotating}
\usepackage{dcolumn}

\newcommand{\Msun}{\mbox{\,M$_\odot$}}

\newcommand{\vunit}{\mbox{\,km\,s$^{-1}$}}
\newcommand{\mic}{\mbox{$\,\mu$m}}
\newcommand{\pion}[2]{{#1}\,{\sc {#2}}}
\newcommand{\fion}[2]{[{#1}\,{\sc {#2}}]}
\newcommand{\ltsimeq}{\raisebox{-0.6ex}{$\,\stackrel
        {\raisebox{-.2ex}{$\textstyle <$}}{\sim}\,$}}
\newcommand{\gtsimeq}{\raisebox{-0.6ex}{$\,\stackrel
        {\raisebox{-.2ex}{$\textstyle >$}}{\sim}\,$}}

\begin{document}
\title[Infrared emission from novae]{Infrared emission from novae} %
\author[A.~Evans \& R. D. Gehrz] {A.~Evans$^1$\thanks{email:
\texttt{a.evans@keele.ac.uk}},  R. D. Gehrz$^2$\thanks{email:
\texttt{gehrz@astro.umn.edu}}\\ $^1$Astrophysics Group, Lennard Jones Laboratory,
Keele University, Keele, Staffordshire, ST5 5BG, UK\\ $^2$Minnesota Institute for Astrophysics, School of Physics \& Astronomy, 116 Church Street SE, \\ University of Minnesota, Minneapolis, MN 55455, USA}

\pubyear{2012}
\volume{00}
\pagerange{\pageref{firstpage}--\pageref{lastpage}}

\date{Received --- ; accepted ---}

\maketitle
\label{firstpage}

\begin{abstract}
We review infrared observations of classical and recurrent novae, at wavelengths
$>3$\mic, including both broad-band and spectroscopic observations.
In recent years infrared spectroscopy in particular has revolutionised our understanding of the nova phenomenon, by revealing fine-structure and coronal lines, and the mineralogy of nova dust. Infrared spectroscopic facilities that are, or will be, becoming available in the next
10--20~years have the potential for a comprehensive study of nova line emission and dust mineralogy, amd for an unbiassed assessment of the extragalactic nova populations.
\end{abstract}


\section{Introduction}\label{s:intro}

In many ways the 1970s marked the beginning of a golden age in our understanding of the
nova phenomenon, as in this decade ultraviolet (UV) and infrared (IR) observations
became available to complement the already well-observed (if not fully understood)
optical observations. Indeed, it can be argued that this pan-chromatic view of the nova
phenomenon was (as in many other areas of astrophysics) the key to understanding the
bigger picture. 

The IR window on novae was opened by the seminal observations by
\cite{geisel}, who obtained broadband photometric observations of Nova
Serpentis 1970 (FH~Ser) from 1--25\mic. The sharp rise in the flux longward of
$\sim2$\mic, coinciding with the deep decline in the visual light curve,
finally confirmed an hypothesis that had been gathering dust (so to speak)
since 1935. In 1934 the nova DQ~Her -- one of the best observed novae of all
time -- erupted and displayed a deep ($\gtsimeq10$~mag) minimum in the visual
light curve (see \cite{martin} for a schematic light curve, based on AAVSO
observations). With extraordinary insight, \cite{mcl} suggested that the deep
minimum might have arisen as a result the formation of dust in the material
ejected in the 1934 eruption, but the technology to make the IR observations
required to confirm the dust hypotheses did not exist at the time.

\begin{figure}
  \centerline{\includegraphics[angle=0,width=8.5cm]{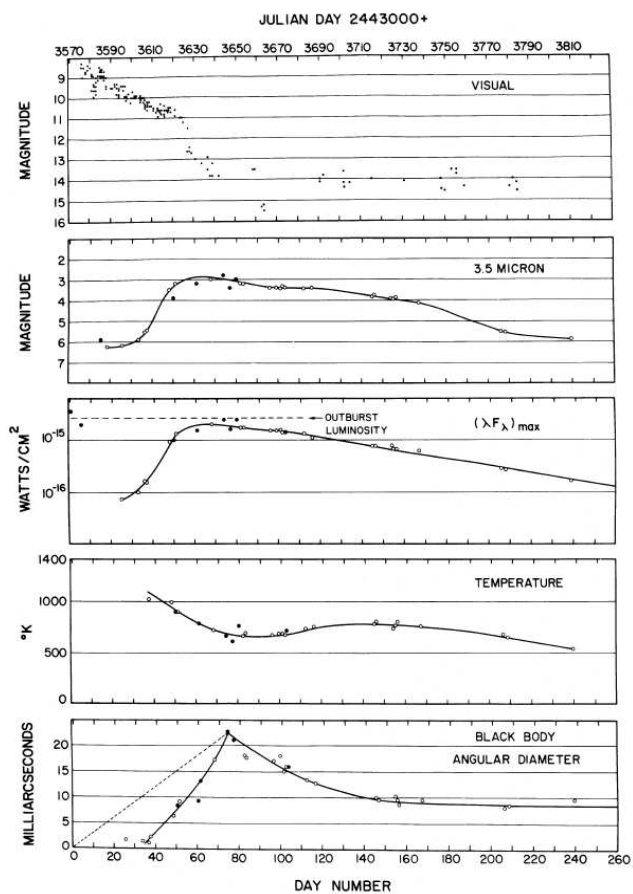}}
  \caption{Optical and IR development of the dusty nova LW~Ser. From top to
  bottom: visual light curve from AAVSO data; 3.5\mic\ light curve; optical-IR
  luminosity, as measured by $\{\lambda{f}_\lambda\}_{\rm max}$, where
  $f_\lambda$ is the flux density in wavelength units; black body temperature of
  the dust; black body angular diameter. See \cite{RDG-lwser} for details.
  \label{qvvul-optir}}
  \end{figure}

The pan-chromatic view of the nova phenomenon has shown that the circumstellar
environment of an erupting nova is extraordinarily hostile, not conducive to the
chemistry required to go from a plasma, via small molecules to dust grains. An
alternative proposal to dust formation was proposed by \cite{BE2}, who suggested that
the dust giving rise to the IR emission pre-dates the eruption, by producing an ``infrared echo''.
However, with one or two exceptions -- in which there is clear evidence for such ``pre-existing''
dust -- it now seems clear that chemistry and dust-formation do indeed proceed in the winds 
of erupting novae. The coincidence of the minimum in the visual light curve and
rise in the IR, as the newly-formed dust reradiates the absorbed visual
light, is well illustrated in the case of the novae NQ Vul \citep{NH} and LW~Ser
\citep[see Fig.~\ref{qvvul-optir};][]{RDG-lwser}. This is now regarded as
irrefutable evidence that the dust forms in the newly-ejected material.  

However there is more to IR observations of novae than the detection and
characterisation of dust. As the eruption proceeds, from the expanding fireball,
through the free-free phase, to the nebular and (in some cases) coronal phases,
the ionisation and excitation of the dispersing ejecta by the still-hot white dwarf results in
a rapidly evolving emission line spectrum. The continuum (free-free) emission
can be used to estimate the ejected mass, while the emission lines can be used to
determine elemental abundances in the ejecta if the state of the still-hot
central engine is known. In the case of fine-structure and coronal emission,
abundances can be in some cases be determined using relatively simple atomic
physics. 

In this contribution we review the IR properties of classical and recurrent
novae at wavelengths longward of $\sim3$\mic; a companion paper in this issue
\citep{DPKB} is devoted to IR observations of novae shortward of 3\mic.


\section{The nova phenomenon: some context}

As is well known \citep[see][for a recent review]{CN2}, a nova explosion occurs
in a semi-detached binary system in which a white dwarf (WD) accretes material
from a cool star via an accretion disc. The base of the accreted material is
compressed and heated, and becomes (electron) degenerate. A Thermonuclear
Runaway (TNR) eventually occurs and, as degeneracy is lifted, the accreted
material is ejected in a nova explosion. Mass transfer then resumes and in time
further nova explosions occur; this may be on a timescale
$\sim10^4\ldots10^5$~years (a classical nova), or on a human timescale,
$\ltsimeq100$~years (a recurrent nova). Indeed, it may be that the recurrence
time is not so markedly bimodal, and that there is a spectrum of recurrence
times, which has not yet become manifest. In any case, whether a classical or a
recurrent, a nova system undergoes many eruptions.

\begin{figure}
  \centerline{\includegraphics[angle=0,width=7cm]{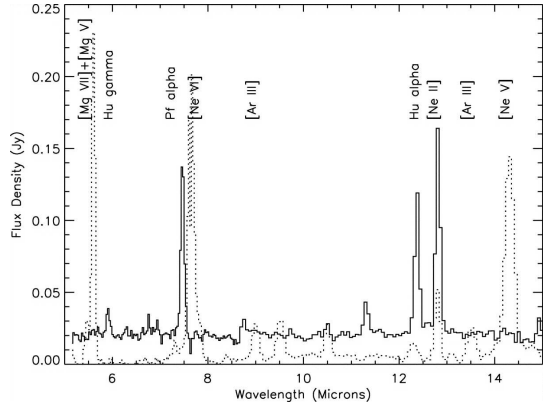}}
  \caption{The contrast between the strength of the neon lines in a CO nova V1186~Sco (solid line) and a neon nova (dotted line) \citep{schwarz-v1186sco}. \label{v1186}}
  \end{figure}

In the course of the explosion some $10^{-5}$ to $10^{-4}$\Msun\ of material is
ejected, at speeds ranging from a few hundred to several thousand \vunit. By
virtue of the fact that the explosion has its origin in a TNR the ejected
material is enriched in CNO (and other elements). Some species observed in nova
ejecta are unlikely to have been generated in the TNR, and it is clear that some
of the WD must be dredged up into the accreted material, as first suggested by
\cite{ferland-cyg-2} to account for the large neon over-abundance in V1500~Cyg.
This led to the realisation that the compact object in the nova binary may be
either be a CO WD or a more massive ONe WD. Nova eruptions on the former tend to
have ejecta masses at the lower end of the mass range and to be the prolific
dust producers, while novae in systems containing ONe WD tend to be faster -- as
the violence of the explosion depends on the WD mass -- eject higher masses, and
seem generally incapable of forming much dust. Nova explosions on ONe WD are often termed
``neon novae'', on account of the great strength of neon lines in their optical
and IR spectra; novae originating on CO WDs are termed ``CO novae''. The
difference in their spectra over the wavelength range $5-15$\mic, where there is a number of neon lines, is well illustrated (see Fig.~\ref{v1186})
in the cases of V1186~Sco (a CO nova) and V1187~Sco (a neon nova). 

The energetics of a nova explosion are well parametrised by the time $t_2$ (or
$t_3$) that the nova's visual light curve takes to decline by 2 (or
3) magnitudes from maxium, although the (extremely) erratic nature of some nova
light curves makes this parameter not only difficult to determine but also to
define. Faster novae (i.e. smaller $t_2$) have greater bolometric luminosities and ejection velocities.
A key point is that fact that the bolometric luminosity remains constant
throughout the early part of the eruption \citep[the TNR ``switches off'' after
about 1\ldots10~years; see][]{krautter} and so as the mass-loss declines, the
effective temperature of the stellar remnant increases as the pseudo\-photosphere
collapses back onto the WD \citep{BH}. This has implications for the evolution of
both the line and dust emission.

\section{Broad-band IR observations} \label{s:broad}

The earliest IR observations consisted of broad-band photometry, most often in
the $J\!H\!K\!L$ (and sometimes $M$) bands\footnote{$J=1.25$\mic, $H=1.65$\mic,
$K=2.2$\mic, $L=3.5$\mic, $M=4.8$\mic}, but also at longer wavelengths. Despite
the limitations of  broad-band observations, they were key to revealing three
important aspects of the nova phenomenon: the formation of dust, the presence of
spectral features in the IR, and the determination of the ejected mass.

As already mentioned, some of the early broad-band observations revealed the
condensation of significant quantities of dust in the ejected material
\citep[e.g.][]{geisel,NH,RDG-lwser}. However even now, more than 40~years on,
the very rapid grain growth in novae remains a poorly understood aspect of the dust
formation process. For example, in the proto-typical heavy dust formers such as
V705~Cas \citep{mason,cas-4} and LW~Ser \citep[see Fig.~\ref{qvvul-optir}][]{RDG-lwser},
the grains grow to maximum size (typically $\sim1$\mic) as judged from the
optical extinction event,
in only 20--40 days. The inference is that grains grow with extremely high
efficiency once they have nucleated. A possible mechanism has been proposed
by \cite{SG}, who argued that photo-ionisation of condensates by the hard
radiation field of the central engine can induce runaway grain growth.
Dust formation may also be facilitated by the clumpiness of nova ejecta,
and the fact that the cores of the clumps are shielded from the hardest
radiation. H-$\alpha$ and thermal IR images of the circum-binary ring around the
hot, luminous, over-contact binary RY~Sct clearly demonstrate that dust can form
behind an ionisation front that has blocked the ionising photons from the dust
formation zone \citep{smithn1,smithn2}.

The nature of the broad-band data meant that little
could be deduced about the nature of the dust, although the expected overabundance
of CNO elements in the ejecta hinted at carbon. As the bolometric luminosity
$L_{\rm bol}$ of the nova remains constant during the early eruption (predicted by
the TNR model of the nova eruption, and confirmed by multi-wavelength
observations) a simple interpretation gives
\[ t_{\rm c} = \left ( \frac{L_{\rm bol}Q_*}{16\pi{v^2}\sigma{T_{\rm c}^{(\beta+4)}}}\right )^{1/2} \]
for the dust condensation time $t_{\rm c}$, in terms of the ejecta speed $v$ and
condensation temperature $T_{\rm c}$. The parameter $\beta$ is defined such that
the IR emissivity of the dust is $\propto\lambda^{-\beta}$, and $Q_*$ is the
absorptivity of the dust, averaged over the spectral energy distribution (SED) of the stellar remnant.

As the dust flows away from the site of the explosion the dust
temperature $T_{\rm d}$ is expected to decline with time $t$ as $T_{\rm
d}\propto t^{-2/(\beta+4)}$ but in some particularly dusty novae (such as NQ~Vul),
the dust temperature remains roughly constant with time \citep[see e.g.][this is well
illustrated in Fig.~\ref{qvvul-optir} for LW~Ser ]{NH,RDG-lwser}
-- the so-called ``isothermal'' stage -- indicating that the
dust physics is rather more complex than indicated above. However as noted by
\cite{ER-pah} the luminosity as seen by the grains is not constant, because of the
way in which ionisation fronts sweep through the ejecta. The isothermal phase has
been interpreted in terms of grain destruction, by chemisputtering of carbon
grains \citep{ME,MEA}, and the photo-processing of carbon dust \citep{ER-pah}.
Indeed, the chemisputtering of amorphous carbon dust by hydrogen led \cite{ME} to
predict the presence of aromatic infrared (AIR)\footnote{The aromatic and
aliphatic hydrocarbon features seen in a wide  variety of astrophysical sources --
including novae -- have been variously referred to as HAC features, PAH features, UIR features
and AIR features; we use the latter designation here.} features in novae,
subsequently discovered in V842~Cen \citep{HMcG} and V705~Cas
\citep[][see Section~\ref{s:air} below]{mason,cas-4}.

Broad-band observations were also sufficient to hint at the presence of the CO
molecule early in the IR development of NQ~Vul \citep[e.g.][]{RDG-nqvul,NH}; the
bandhead of the CO fundamental vibrational transition is at 4.6\mic, which
elevates the $M-$band flux. Indeed the data presented by \cite{RDG-nqvul}
clearly show that lines (probably the CO bandhead) contributing to the 5\mic\
excess were present before the dust formed. This molecule is key to
understanding the chemistry that leads to dust formation. While the conventional
paradigm is that the C:O ratio (by number) in  the ejected material determines
the nature of the dust formed, this assumes that CO formation goes to saturation
so that whichever of C and O has the lower abundance is locked up in CO; this
seems not to apply to nova ejecta, in which CO formation does not go anywhere
near saturation, giving rise to the production of a variety of dust types. An
alternative explanation for the latter is that there are strong abundance
gradients within the ejecta. 

Early broad-band observations also hinted at the presence of the
\fion{Ne}{ii}12.81\mic\ fine-structure line, in the IR observations of V1500~Cyg
\citep{ennis,ferland-cyg}, giving rise (as noted above) to the neon nova concept.
As we discuss below, the IR fine-structure (and coronal) lines provide a means of
estimating ejecta temperature and abundances. 

The mass ejected ($M_{\rm ej}$) in a classical nova eruption is a key parameter.
It is central to the understanding of the TNR, to the determination of abundances,
and to the contribution that novae make to the chemical evolution of the
Galaxy \citep[see e.g.][and chapters and references therein, for a full discussion
of ejected masses]{CN2}. It is also key to the fate of the WD in the nova system.
If $M_{\rm ej}$ is less than the mass accreted by the WD then the mass of the
latter must increase with time, with implications for a potential link between
nova systems and the progenitors of Type~Ia supernovae; but if $M_{\rm ej}$
exceeds the accreted mass then novae are likley greater contributors to Galactic
chemical evolution than hitherto thought. Either way this one parameter lends novae a great
significance in the bigger picture.

The ejected mass may be determined by multi-wavelength broad-band IR
observations during the free-free phase; the observations fix the wavelength
$\lambda_{\rm c}$ at which the free-free emission becomes optically thick,
a measure of the shell density
\citep[see][and references therein]{RDG-CN2}.  Masses can also be determined
using a Thomson scattering model when IR observatons show that the fireball is
becoming optically thin \citep{RDG-CN2}. Ejected masses determined by these two
methods agree when applied to the same nova and are   typically in the range 0.5
$-10\times10^{-5}$\Msun\ \citep[see][for a summary]{RDG-pasp, RDG-CN2}; ejected
masses determined from radio observations are comparable \citep[see][for a
summary]{ERS}, although IR and radio observations sample different regions of
the ejected material \citep{RDG-CN2}. These observationally-determined masses
are significantly greater than those expected from the TNR theory
\citep[e.g.][]{starrfield-CN2}, and the discrepancy is all the greater when it
is noted that neither IR nor radio observations include any neutral material.

\section{IR spectroscopy} \label{irspec}

IR spectroscopy (with spectral resolution $R = \lambda/\Delta\lambda\gtsimeq50$)
became widespread in the 1980s and it is at this time that IR observations of
novae began to rival optical observations in their diagnostic capability. In
particular spectroscopic observations from observatories in space, especially
the {\it Infrared Space Observatory} \citep[{\it ISO};][]{iso} and more recently
the {\it Spitzer Space Telescope} \citep{spitzer,RDG-spitzer}, have been pivotal
in advancing our understanding of the circumstellar environments of novae.

\subsection{Line emission}

As is well known, when the visual light curve has declined by about
3~magnitudes, the optical spectrum of a classical nova makes the dramatic
``transition'' from absorption to emission \citep[see][for a very nice
illustration of this in the case of DQ~Her]{PG}. Much else happens around this
time, such as the formation of dust in the ejected material. Thereafter, the
ionisation states of the emission lines increase as the pseudo\-photosphere
collapses onto the white dwarf (see above), with specific ionisation states
appearing at well-defined epochs in the evolution of the eruption
\citep[see][especially Figure~5, for an early discussion of this point]{bath}.
As the nova eruption progresses, a number of fine-structure lines appear
longward of 3\mic\ and, in some cases, coronal lines appear as well.

\subsubsection{The nebular phase} \label{sss:nebular}

  \begin{figure}
  \centerline{\includegraphics[angle=0,width=10cm]{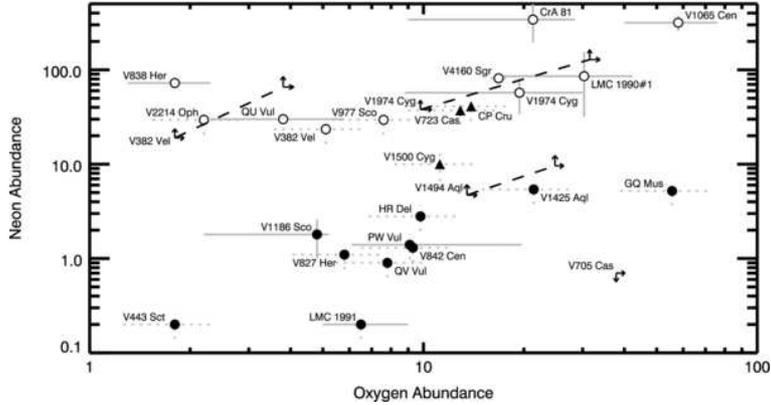}}
  \caption{Neon over-abundances versus oxygen over-abundances (in both cases
  relative to solar) for a variety of recent novae.
  The open circles represent the ONe targets, the filled circles represent the
  CO novae, and the filled triangles are atypical novae. It is clear that the
  strong dust formers fall in the regime where the neon abundance is 1--5 times
  solar, and that the neon novae have neon abundances that are much higher than solar.   
  From \cite{helton-spitzer}; see this paper for further
  details.\label{abundance-summary-spit}}
  \end{figure}

The evolution of the nebular spectrum of novae is well observed and hydrogen
recombination lines appear very early in the development of the nebular spectra;
prominent series with lines in the wavelength range of the {\it Spitzer} 
Infrared Spectrograph \citep[IRS;][]{irs}
include the Humphreys ($n\rightarrow6$, $\lambda_\alpha=12.37\mic$),
$n\rightarrow7$ ($\lambda_\alpha=19.06\mic$) and $n\rightarrow8$
($\lambda_\alpha=27.80\mic$) series, where $n$ is the principal quantum number
of the upper level.

The nebular phase offers the potential to determine elemental abundances in the
ejected material (and hence of the material produced in the TNR). However the
determination of abundances can be fraught with difficulties, including (a)~the
poorly known interstellar extinction (and its wavelength-dependence) to novae in
general, (b)~the possible wavelength dependence of any circumstellar extinction
\citep[but in the case of V705~Cas this was known to be neutral, at least in the
UV;][]{shore}, (c)~the clumpiness of nova ejecta, and (d)~the physical state of
the ionising and/or exciting mechanism. In general the ejected material is in
general likely to be highly asymmetric (e.g. displaying equatorial rings and
polar caps) and to have abundance gradients.

However, longward of 3\mic\ the reddening becomes a lesser issue, and
observations with the {\it Spitzer} IRS
have played a prominent role in the determination of elemental abundances in
nova ejecta. Determinations of nebular abundances generally make use of
well-tested photo\-ionisation codes such as {\sc cloudy} \citep[][and earlier 
papers]{cloudy}, which necessarily make simplifying assumptions, e.g. regarding
the geometry of the ionised material, although {\sc nebu} \citep{morriset}
allows for a range of nebular geometries.
Given the prominence of hydrogen recombination (and other) lines it is
relatively straight-forward to express abundances relative to H, and then
compare this with an assumed solar abundance\footnote{Care should be exercised
in comparing nova abundances relative to solar as the latter are occasionally
revisited; see \cite{shore-CN2} for a discussion of this point.}
\citep[e.g.][]{asplund}.

  \begin{figure}
  \centerline{\includegraphics[angle=0,width=6cm]{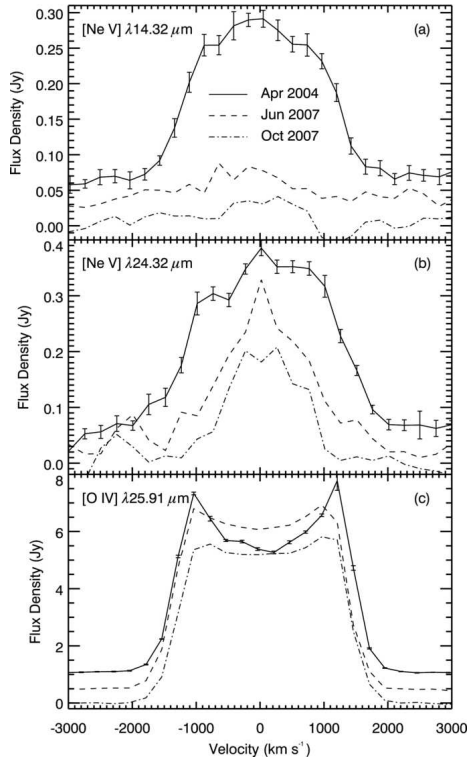}}
  \caption{Line profiles for three fine-structure lines in the IR spectrum of
  V1494~Aql. Note the difference between the profiles of the neon and oxygen
  lines. From \cite{helton-spitzer}. \label{profiles_1}}
  \end{figure}

A summary of abundances in nova ejecta,
including some values obtained with {\it ISO} and {\it Spitzer},
is given by \cite{RDG-CN2}, while compilations of abundances
deduced using {\it ISO} and {\it Spitzer} data (as well as data at shorter
wavelengths) are given by \cite{shore-CN2} and \cite{helton-spitzer}. Caution
should be exercised in comparing these abundances as they are expressed relative
to differently-assumed solar abundances. 

In Fig.~\ref{abundance-summary-spit}, we show the relative abundances of neon
and oxygen in nova ejecta, as determined from {\it ISO} and {\it Spitzer}
observations, together with observations at shorter wavelengths
\citep{helton-spitzer}; note that the Ne abundance for QU~Vul in this Figure is
from \cite{schwarz-qu} rather than the more recent determination by
\cite{RDG-quvul}. Also the abundances in this figure have been adjusted to a
standard solar value). Note that the novae in this figure do not form a
homogeneous sample: they are generally selected on the grounds that they happened
to trigger a target-of-opportunity programme. However there is a clear
demarcation between CO and neon novae, the latter having neon overabundance
relative to solar of $\gtsimeq20$.

Line profiles during the nebular phase have long been known to be complex and
this is also the case for IR emission lines. A selection of line profiles for
V1494~Aql is shown in Fig.~\ref{profiles_1} \citep[from][]{helton-spitzer}. The
saddle-shaped profile of the \fion{O}{iv}25.91\mic\ line resembles that of the
optical emission lines seen early in the outburst, and indicates an equatorial
cap/polar cap structure in the ejecta. The profiles of the neon lines, on the
other hand, are less complex, and may indicate that the emission arises in a
geometrically thick, but optically thin, shell \citep{helton-spitzer}. Such
differences may indicate that these ions reside in different regions of the
ejecta (this may also be suggested by the range of dust types formed in nova
ejecta; see Section~\ref{s:dust} below).

\subsubsection{Fine structure emission} \label{FSE}

The ground states of heavy elements may have fine structure having the same $L$ and $S$ quantum numbers but different $J$. The upper level is easily populated
(usually by electron collision) and radiative de-excitation may then occur giving
rise to fine structure line emission; the transitions are highly forbidden, and
the upper level is collisionally de-exited if the electron density exceeds a
specific value $n_{\rm crit}$. Such transitions may act as extremely efficient
coolants in low density nebulae \citep[see][who made an early prediction of the
likely strength of \fion{O}{iv}25.89\mic\ and other far-IR fine-structure
lines in novae]{ferland-dq}.

 \begin{figure}
  \centerline{\includegraphics[angle=0,width=5.cm]{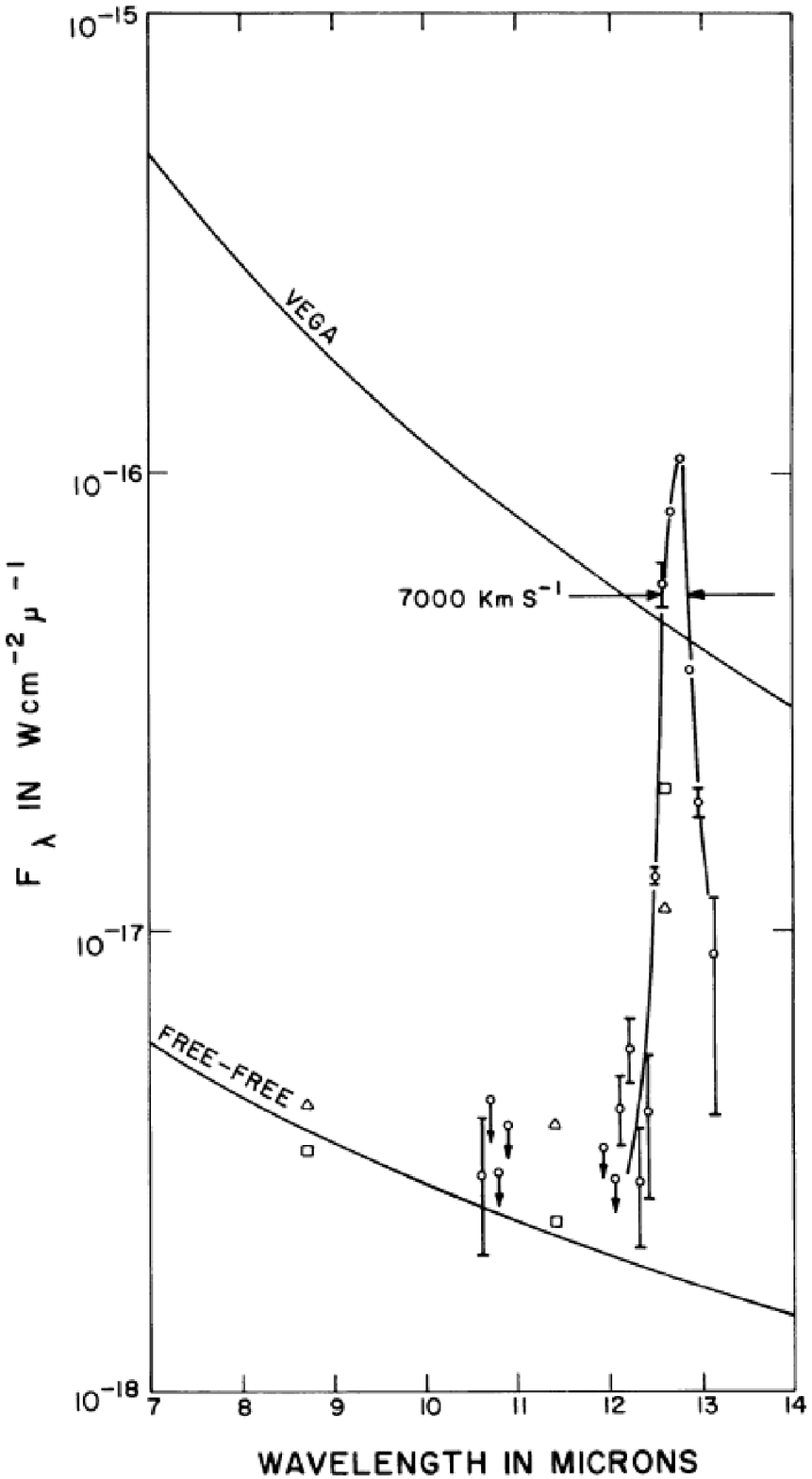} \qquad
               \includegraphics[angle=0,width=6.cm]{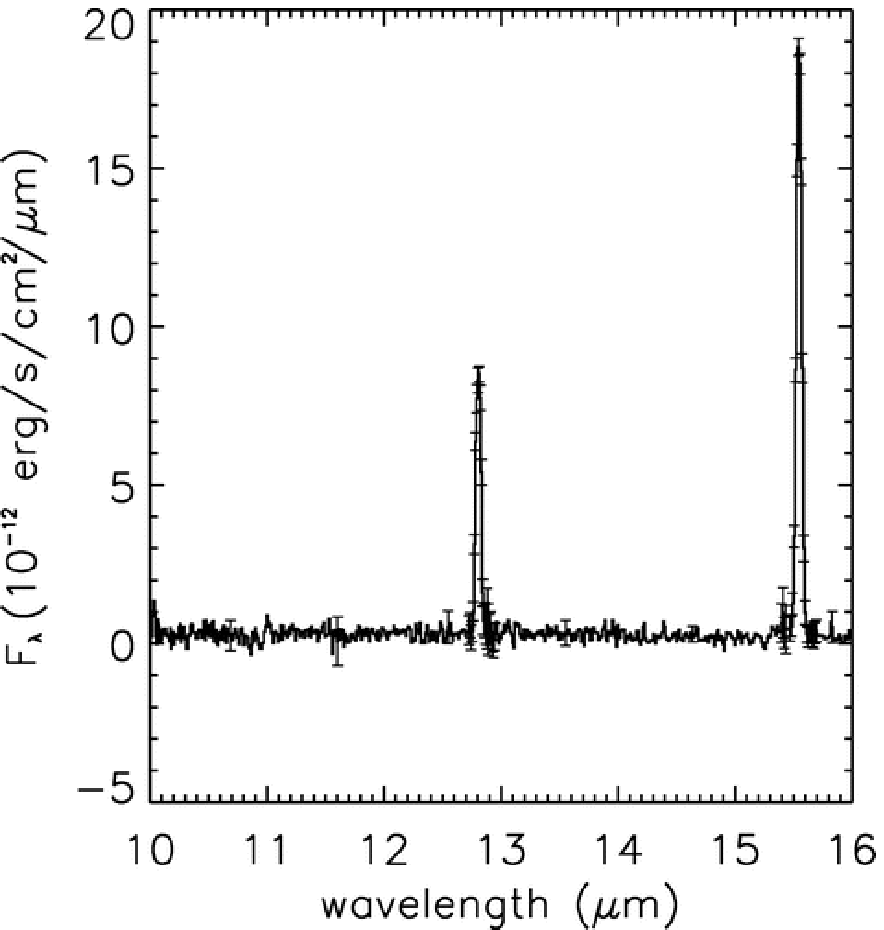}}
  \caption{Left: Ground-based IR spectrum of QU~Vul in the $8-13$\mic\ window,
  showing free-free emission and strong emission in the \fion{Ne}{ii}12.81\mic\
  fine-structure line. From \cite{RDG-quvul-1}. Right: Spectrum of QU~Vul in the
  same wavelength range using the {\it Spitzer} IRS \citep{RDG-quvul}. 
  \label{quvul}} 
  \end{figure}

\fion{Ne}{ii} was the first IR fine-strucure line to be observed in a nova
(V1500~Cyg) and on this basis \cite{ferland-cyg} suggested that both
\fion{Ne}{ii}15.56\mic\ and \fion{O}{iv}25.89\mic\ would also be strong. QU~Vul
(Fig.~\ref{quvul}) was subsequently discovered to have the strongest
\fion{Ne}{ii}12.8\mic\ relative to the continuum ever observed in an
astrophysical source \citep{RDG-quvul-1}, thus becoming the ``quintessential''
example of a neon nova. 
\fion{O}{iv} was subsequently found to be very prominent in the spectra of V1974~Cyg
\citep{salama-1974}, V705~Cas \citep{salama-cas} and V1425~Aql \citep{lyke-cpcru}
in data obtained with {\it ISO}; see also below. More recently observations of novae with the {\it Spitzer} IRS have also found prominent \fion{O}{iv}, and also
\fion{Ne}{ii}12.81\mic, \fion{Ne}{iii}15.55\mic, \fion{Ar}{iii}8.99\mic\ and
\fion{S}{iv}10.51\mic\ (see Fig.~\ref{quvul}).

The importance of these lines is that they may provide a
relatively elementary estimate of the abundance of the emitting ion. The ion may be
treated as a simple two-level system and detailed balance
then requires that the total number of emitting ions $N$ is
\[ N = \frac{L\lambda}{n_{\rm e}hc} \:\: \frac{1 + [n_{\rm e}/n_{\rm
crit}]}{q_{12}} \] 
where $\lambda$ is the wavelength of the transition, $L$ is the luminosity in the
transition and $q_{12}$ is the (temperature-dependent) collisional excitation rate
\citep{RDG-quvul}. In the low density limit (i.e. $n_{\rm e} \ll n_{\rm crit}$),
every collisional excitation is followed by a radiative de-excitation.

This technique has been used by \cite{RDG-quvul} to determine a lower limit on
the neon abundance in the ejecta of the neon nova QU~Vul. Fig.~\ref{quvul} shows
the spectrum of this nova, obtained from the ground in the $8-13$\mic\ window
in 1984, 140~days after outburst \citep{RDG-quvul-1}, and in 2004 -- over a
similar wavelength range -- with the {\it Spitzer} IRS. Strong emission
in the \fion{Ne}{ii} fine structure line is evident, as is (in 2004) emission by
\fion{Ne}{iii}15.55\mic. \cite{RDG-quvul} concluded that QU~Vul was
over-abundant in neon (relative to solar) by a factor of at least 168 some
19~years after eruption. Such large overabundances, combined with the observed
ejected masses, suggest that novae are not insignificant contributors to the
chemical evolution of the Galaxy.

A list of fine-structure lines longward of 3\mic\ expected to be observed in
novae is given in Table~\ref{FS}, in which E.P and I.P. are respectively the
excitation potential of the upper level and the ionisation potential needed to
form the ion respectively. The list includes a number of lines that are
commonly seen in the IR spectra of novae, such as \fion{Ca}{iv}3.20\mic\ and
\fion{Ne}{ii}12.81\mic. Also included are various sodium lines, such as
\fion{Na}{iii}7.31\mic. $^{22}$Na is predicted to be produced in the TNR,
particularly on high-mass WDs. This isotope is radioactive and undergoes
$\beta^+-$decay to $^{22}$Ne with half-life 2.6027~years; this decay should be
observable as the line spectrum evolves. However, while a number of neon lines
are well-observed in the IR spectra of novae, the sodium lines have so far
proved elusive.

\begin{table}
  \caption{Selected fine-structure lines with wavelength $>3$\mic, from \cite{fs-list}.}\label{FS}
  \medskip
  \begin{center}
    \begin{tabular}{lcrrr}\hline      
    \multicolumn{1}{l}{Species} &   \multicolumn{1}{c}{Transition} & \multicolumn{1}{c}{$\lambda$ ($\mu$m)} & E.P. (eV) & I. P. (eV) \\\hline
\fion{Ca}{iv}    &   $^2$P$_{1/2}-^2$P$_{3/2}$   & 3.2067 & 50.91 & 67.27 \\
\fion{Ca}{v}     & $^3$P$^1-^3$P$^2$ &              4.1594 & 67.27  & 84.50 \\
\fion{Ar}{vi}    &  $^2$P$_{3/2}-^2$P$_{1/2}$   &              4.5295 & 75.02  & 91.01 \\
\fion{Na}{iii}   &   $^2$P$_{1/2} - ^2$P$_{3/2}$  &              7.3177 &   47.29  &71.62 \\
\fion{Ar}{v}     &    $^3$P$_2 - ^3$P$_1$     &             7.9016   &59.81  &75.02\\
\fion{Ar}{iii}   &   $^3$P$_1 - ^3$P$_2$      &           8.9914    &  27.63 & 40.74 \\
\fion{Na}{iv}    &    $^3$P$_1 - ^3$P$_2$     &           9.0410     &71.62  &98.91\\
\fion{S}{iv}    &     $^2$P$_{3/2} - ^2$P$_{1/2}$  &         10.5105    & 34.79  &47.22 \\
\fion{Ca}{v}    &     $^3$P$_0 - ^3$P$_1$      &          11.4820   & 67.27  & 84.50 \\
\fion{Ne}{ii}   &   $^2$P$_{1/2} - ^2$P$_{3/2}$    &         12.8136    & 21.56  & 40.96\\
\fion{Ar}{v}    &     $^3$P$_1 - ^3$P$_0$     &          13.1022    &  59.81  & 75.02\\
\fion{Ne}{iii}  &   $^3$P$_1 - ^3$P$_2$        &          15.5551     & 40.96  & 63.45\\
\fion{S}{iii}   &    $^3$P$_2 - ^3$P$_1$      &           18.7130    & 23.34  & 34.79\\
\fion{Na}{iv}   &    $^3$P$_0 - ^3$P$_1$       &        21.2900      & 71.62   & 98.91\\
\fion{Ar}{iii}  &   $^3$P$_0 - ^3$P$_1$       &          21.8302      & 27.63  & 40.74\\
\fion{O}{iv}   &   $^2$P$_{3/2} - ^2$P$_{1/2}$   &        25.8903      & 54.93  &77.41\\
\fion{S}{iii}   &    $^3$P$_1 - ^3$P$_0$       &       33.4810        & 23.34  &34.79\\
\fion{Ne}{iii}  &   $^3$P$_0 - ^3$P$_1$       &         36.0135      &  40.96 & 63.45\\
\fion{O}{iii}  &    $^3$P$_2 - ^3$P$_1$        &       51.8145        & 35.12 & 54.93\\
\fion{N}{iii}  &  $^2$P$_{3/2}-^2$P$_{1/2}$    &         57.3170      &  29.60  & 47.45\\
\fion{O}{i}     &    $^3$P$_1-^3$P$_2$      &        63.1837       &  0.00 & 13.62\\
\fion{O}{iii}   &   $^3$P$_1-^3$P$_0$         &        88.3560     &   35.12 & 54.93 \\
  \hline
    \end{tabular}\\[5pt]
  \end{center}
\end{table}

\begin{figure}
  \centerline{\includegraphics[angle=-90,width=10cm]{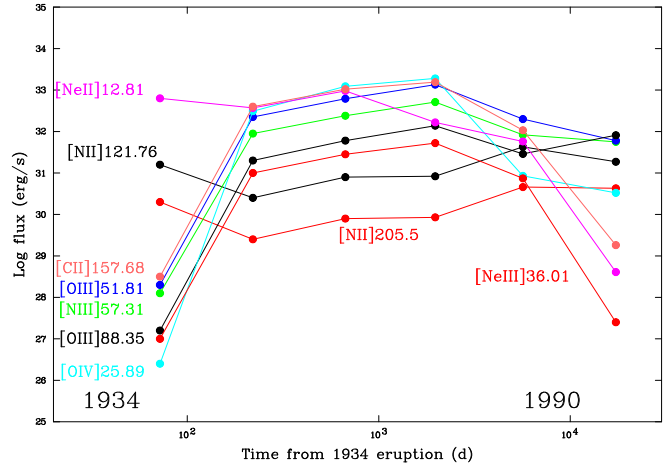}}
  \caption{Predicted variation in fine structure line fluxes, over a period
  $\sim70$~years, for DQ~Her 1934. Based on   \cite{martin2}. \label{MARTIN}}
  \end{figure}

In a careful and thorough discussion of the evolution of the emission line
spectrum of DQ~Her, \cite{martin,martin2} has predicted the variation of the
emission line fluxes from $\sim40$~days to $\sim50$~years after eruption, for
emission lines in the UV, optical and IR/far-IR. With the caveat that
this calculation is for the elemental abundances, physical conditions and the
stellar remnant of DQ~Her, the expected variation with time is shown in
Fig.~\ref{MARTIN}. Note that some lines, such as
\fion{Ne}{iii}36.01\mic, decline significantly with time; others, such as
\fion{O}{iv}25.89\mic, are expected to remain strong for many years, even
decades, after outburst. The observed strength of the \fion{O}{iv} line is borne out by {\it ISO} and {\it Spitzer} observations of V1974~Cyg \citep{salama-1974,helton-spitzer},
V705~Cas \citep{salama-cas}, V1425~Aql \citep{lyke-v1425}, CP~Cru \citep{lyke-cpcru},
V1187~Sco \citep{lynch-1187}, QU~Vul \citep{RDG-quvul}, V2362~Cyg \citep{lynch-2362},
DZ~Cru \citep{dzcru}, V1065~Cen \citep{helton-v1065}, V328~Vel and V1494~Aql
\citep{helton-spitzer}, and in the recurrent nova RS~Oph \citep{rsoph-1,rsoph-2}

\subsubsection{Coronal emission} \label{s:coronal}

A working definition of coronal emission is given by \cite{greenhouse} as
emission lines ``arising from ground-state fine-structure transitions in
species with ionisation potential (I.P.) $>100$~eV''. The distinction from fine
structure lines is in the high energy needed to form the ions concerned.
From the nova perspective, coronal emission was first observed optically
in the recurrent nova RS~Oph \citep[see][for a retrospective]{wallerstein}.
Coronal lines are often observed in the IR spectra of novae, but
not all novae undergo a coronal phase.

The high ionisation states observed in coronal emission may arise from
photo\-ionisation \citep[e.g. novae are known X-ray emitters during and after
eruption,][]{krautter}. Alternatively, they may be excited by collisional
ionisation if there are strong shocks
in the nova environment, either between ejecta moving at different speeds or by
the interaction of ejected material with a pre-existing wind. The required
strong interaction between ejecta and a stellar wind is well established in the
case of the recurrent nova RS~Oph (see Section~\ref{s:recurrent} below).
A list of actual and potential coronal lines longward of 3\mic\ that may be
present in the IR spectra of novae is given in Table~\ref{t:coronal}.

A discussion of IR coronal line emission in novae is given by \cite{greenhouse}.
These authors use coronal line intensity ratios to determine that the electron
temperatures in the coronal zones in novae are $\sim3\times10^5$~K. Abundances
are then estimated from individual line intensities and by summing over
ionisation states; this latter step has to assume a specific form for the
collisional ionisation equilibrium as a function of temperature. As noted by
\cite{greenhouse} the outcome is very dependent on the plasma model assumed.

\begin{table}
  \caption{Selected coronal lines \citep[as defined by][]{greenhouse} with wavelength $>3$\mic,
  from \cite{fs-list}. See text for definition of ``coronal''.}\label{t:coronal}
  \medskip
  \begin{center}
    \begin{tabular}{lcrrr}\hline      
    \multicolumn{1}{l}{Species} &   \multicolumn{1}{c}{Transition} & \multicolumn{1}{c}{$\lambda$ ($\mu$m)} & E.P. (eV) & I. P. (eV) \\\hline
\fion{Mg}{viii}  & $^2$P$^{3/2} - ^2$P$^{1/2}$ & 3.0279 & 224.95 & 265.96 \\
\fion{Al}{vi}    &  $^3$P$_1-^3$P$_2$       &    3.6597 & 153.83 & 190.48 \\
\fion{Al}{viii}  & $^3$P$_2-^3$P$_1$        &              3.6900 & 241.44 & 284.60 \\
\fion{Si}{ix}    &   $^3$P$_1-^3$P$_0$      &              3.9357 & 303.17 & 351.10 \\
\fion{Ca}{vii}   &   $^3$P$_2-^3$P$_1$      &              4.0858 & 108.78 & 127.20 \\
\fion{Mg}{iv}    &  $^2$P$_{1/2} - ^2$P$_{3/2}$  &   4.4867 & 80.14 & 109.24 \\
\fion{Na}{vii}   &   $^2$P$_{3/2}-^2$P$_{1/2}$  &              4.6847 & 172.15 & 208.44 \\
\fion{Mg}{vii}   &   $^3$P$_2-^3$P$_1$      &              5.5032 & 186.51 & 224.95\\
\fion{Mg}{v}     & $^3$P$_1-^3$P$_2$        &              5.6099 & 109.24 & 141.27\\
\fion{Al}{viii}  &  $^3$P$_1-^3$P$_0$       &              5.8500 & 241.44 & 284.60\\
\fion{Ca}{vii}    &   $^3$P$_1-^3$P$_0$      &              6.1540 & 108.78 & 127.20\\
\fion{Si}{vii}   &   $^3$P$_0-^3$P$_1$      &              6.4922 &  205.05 &  246.52\\
\fion{Ne}{vi}    &  $^2$P$_{3/2} - ^2$P$_{1/2}$    &             7.6524  &  126.21&  157.93\\
\fion{Na}{vi}    &    $^3$P$_2-^3$P$_1$     &            8.6106    &  138.39 & 172.15\\
\fion{Mg}{vii}   &   $^3$P$_1-^3$P$_0$      &            9.0090   & 186.51 & 224.95\\
\fion{Al}{vi}   &    $^3$P$_0-^3$$_1$     &            9.1160     &  153.83 & 190.48 \\
\fion{Mg}{v}   &     $^3$$_0-^3$P$_1$      &          13.5213    & 109.24  & 141.27\\
\fion{Ne}{v}  &   $^3$P$_2 - ^3$P$_1$           &         14.3217    & 97.12 & 126.21\\
\fion{Na}{vi} &   $^3$P$_1-^3$P$_0$      &             14.3964    & 138.39 & 172.15\\
\fion{Ne}{v}    &   $^3$P$_1 - ^3$P$_0$         &        24.3175        & 97.12 & 126.21\\
  \hline
    \end{tabular}\\[5pt]
  \end{center}
\end{table}

\subsection{Dust emission} \label{s:dust}

As noted in Section~\ref{s:intro} the idea of dust formation in nova winds has
been around since the 1930s and it is now a well-observed phenomenon in the
IR. A review of dust formation in novae is given by \cite{ER-CN2} and
\cite{RDG-CN2}; see also \cite{shore-CN2}.

There have been several attempts to correlate the dust-forming capabilities of
novae with other parameters, such as speed class, outburst luminosity, ejected
mass, and outflow velocity; see \cite{GNey}, \cite{RDG-araa} and \cite{ER-CN2}
for discussions. However it is likely that whether or not a nova forms dust
depends on the complex interaction between a number of parameters, but it is
apparent from Fig.~\ref{abundance-summary-spit} that there is an
anti-correlation between the presence of strong coronal lines and the nova's 
ability to form dust.

While the early (broad-band) IR observations found evidence for dust, they were not
sufficiently detailed to say anything about the composition of the dust
(although carbon was a strong candidate). However IR spectroscopy has been able
to reveal spectral features of dust, including silicates and hydrocarbons.

\subsubsection{Nature of the dust} \label{ss:nature}

Several cases exist showing that multiple types of astrophysical grains can form
in the ejecta of a single nova. V842~Cen is known to have formed amorphous
carbon, silicates, and hydrocarbons \citep[see][]{RDG-madrid}, while QU~Vul formed
amorphous carbon, SiC, silicates, and hydrocarbons as the ejecta evolved
\citep{RDG-quvul-0}.

\begin{figure}
\centerline{\includegraphics[angle=0,width=12cm]{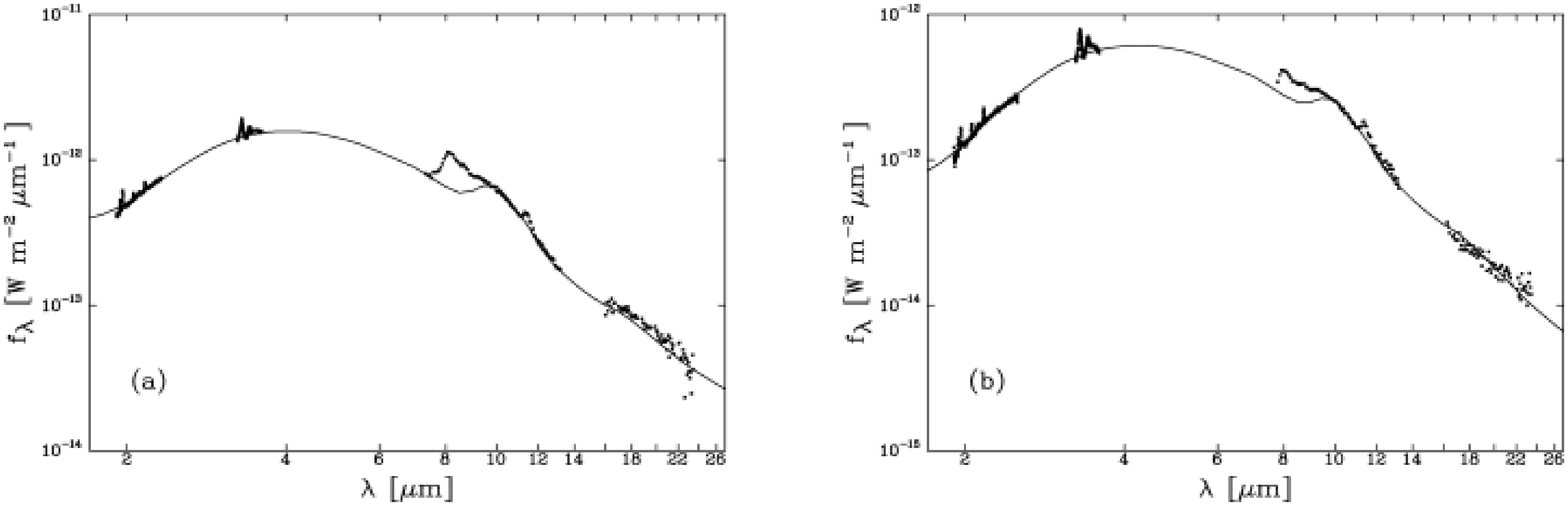}} \medskip
\centerline{\includegraphics[angle=0,width=12cm]{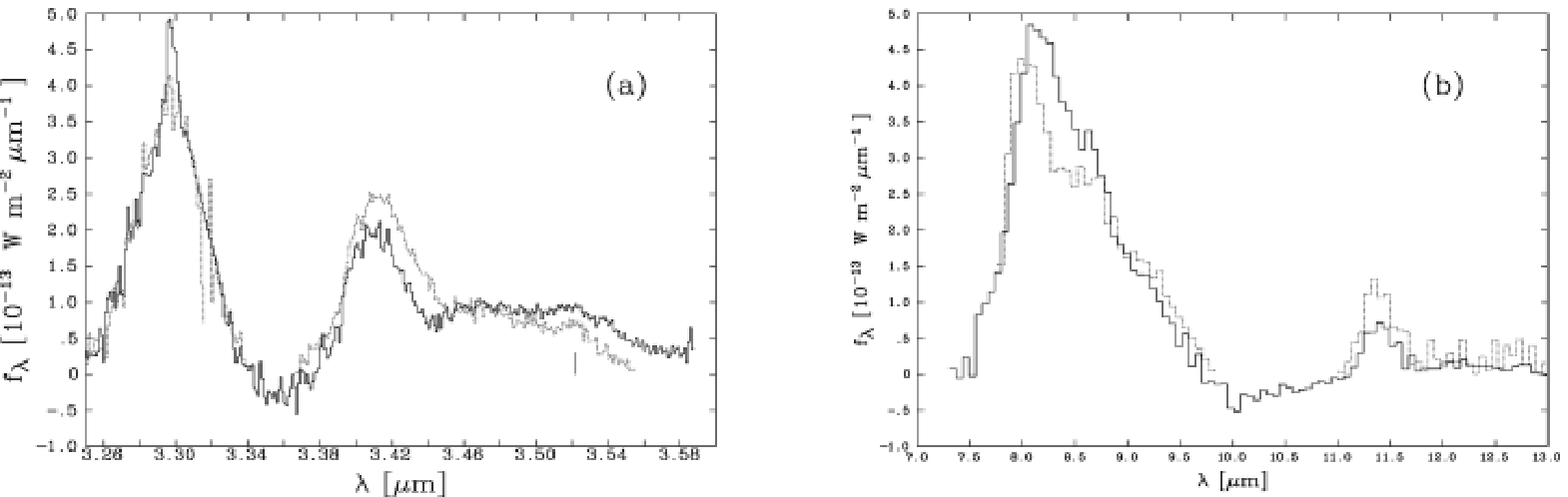}}
  \caption{Ground-based IR spectra of V705~Cas, data obtained 253 (left) and 320
  (right) days after outburst. Top: SED from $2-20$\mic; curves are fits using
  the {\sc dusty} code \citep{dusty}. 
  Bottom: close-up of the (continuum-subtracted) 3.28\mic\ and 3.4\mic\ AIR
  features, which seem like ``noise'' in the SEDs in the upper frames. From
  \cite{cas-4}. \label{cas93}}
\end{figure}

V705~Cas was a particularly well-observed dusty nova
\citep{mason,cas-93,cas-4}, and displayed the characteristic deep minimum in the
visual light curve. Ground-based IR spectroscopy was obtained for this object, from
$1-20$\mic\ (Fig.~\ref{cas93}). The dust in this nova typifies many of the
unusual (even anomalous)
  \begin{figure}
  \centerline{\includegraphics[angle=0,width=9cm]{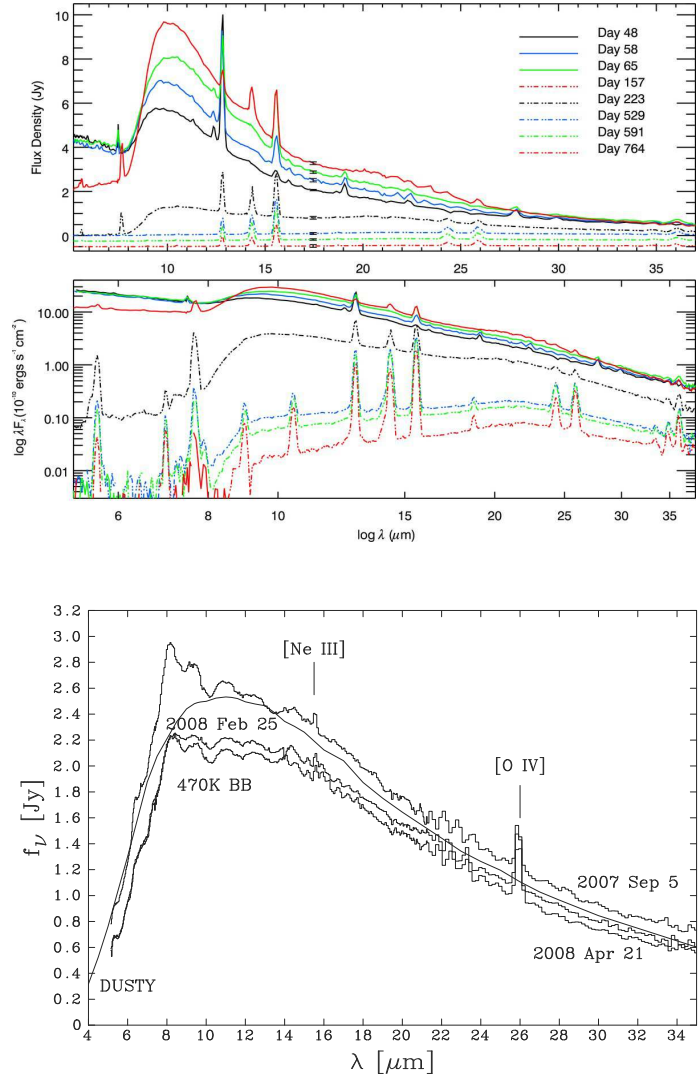}}
  \medskip 
  \centerline{\includegraphics[origin=lt,angle=-90,width=10cm]{dzcru_dust.eps}}
  \caption{Top: Evolving {\it Spitzer} spectrum of V1065~Cen, showing a
  prominent 9.7\mic\ silicate feature \citep{helton-v1065}. Bottom: Evolving
  {\it Spitzer} spectrum of DZ~Cru, showing strong emission by carbon dust, and
  the presence of AIR features at 6.45\mic, 7.24\mic, 8.12\mic, 9.29\mic,       
  10.97\mic\ and 12.36\mic\ \citep{dzcru}. \label{v1065-cen}} 
  \end{figure}
facets of nova dust, including the condensation of both oxygen- and carbon-rich
dust, and the presence of AIR features. The presence of the silicate dust
signature at 9.7\mic\ -- normally indicative of an oxygen-rich environment for
grain formation -- and carbon/AIR emission -- indicative of a carbon-rich
environment -- is at first sight paradoxical and contrary to the ``CO paradigm''
(see Section~\ref{s:broad}). The paradox may be resolved by the possibility
that, in nova winds, the pre-dust chemistry does not enable CO formation to
proceed to saturation, thus allowing the formation of both oxygen- and
carbon-rich grain types. Alternatively, there may be significant abundance
gradients in the ejected material, for example between the polar plumes and the
equatorial ejecta in systems where rotation is high \citep[see][]{RDG-quvul-0}.
Such gradients are also suggested by emission line profiles (see
Section~\ref{sss:nebular} above).

  \begin{figure}
  \centerline{\includegraphics[angle=270,width=7cm]{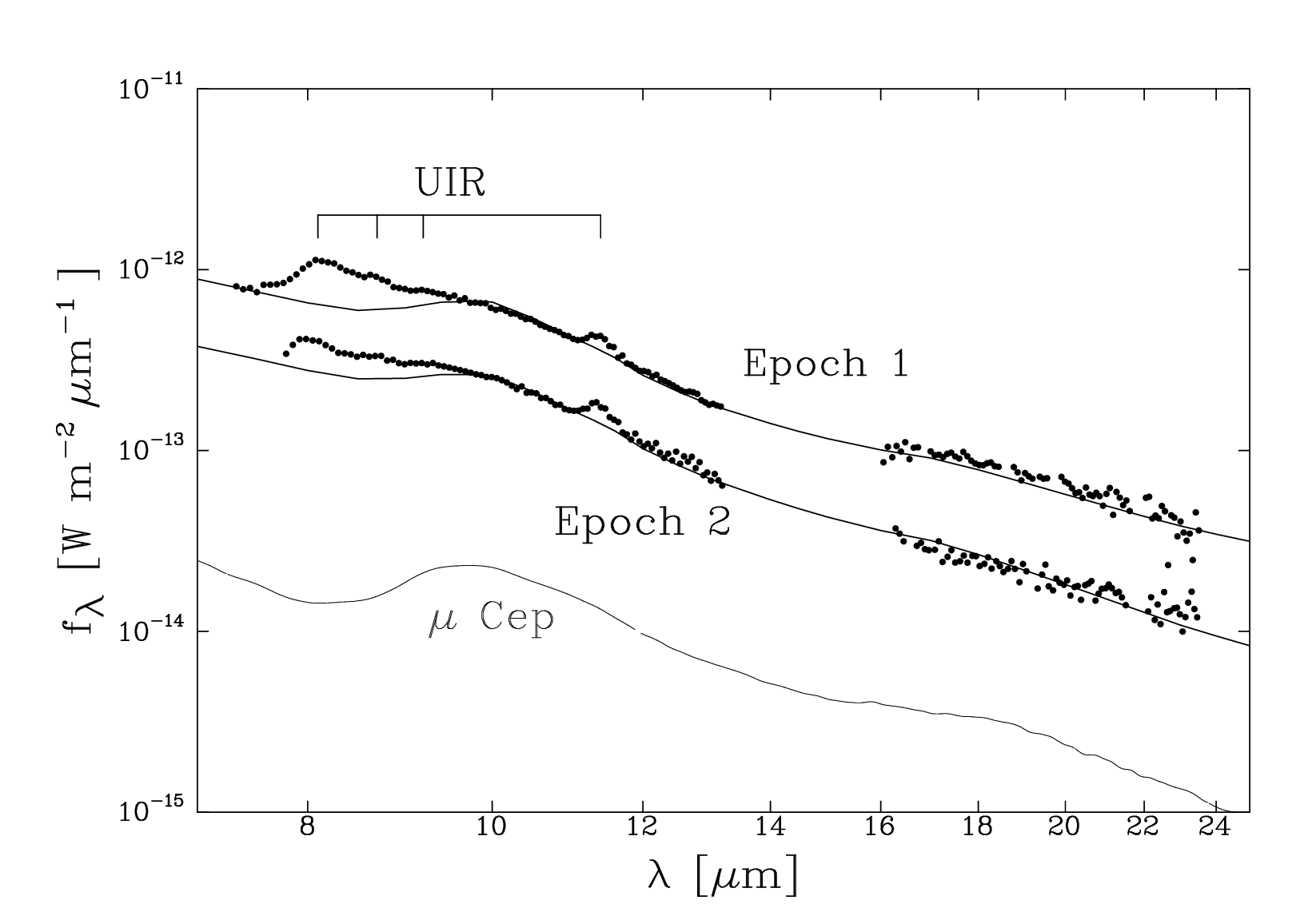} \quad
       \includegraphics[angle=270,width=7cm]{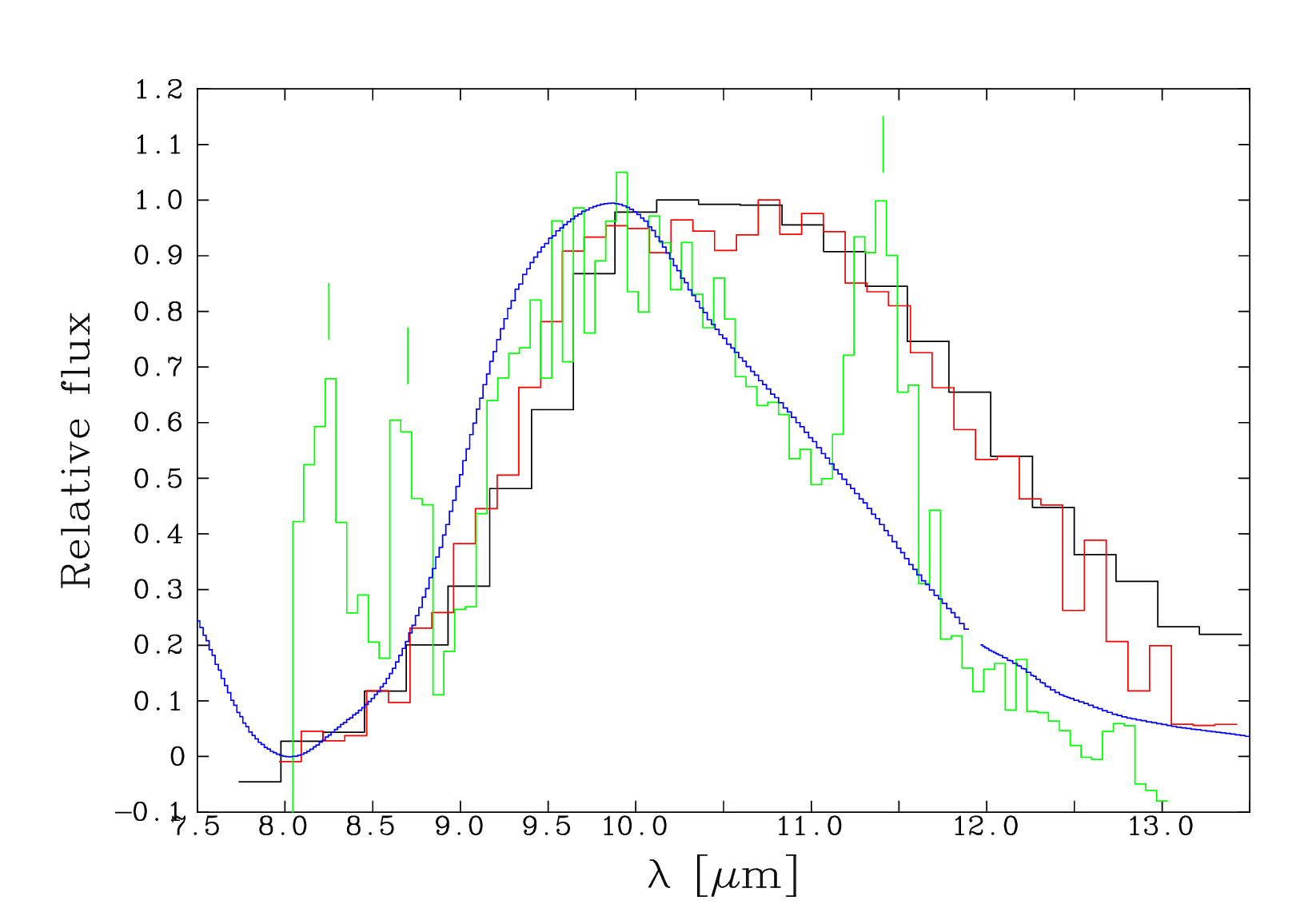}} 
  \caption{Left: The 9.7\mic\ silicate feature in V705~Cas; note the weak
  18\mic\ feature. The silicate feature in the red supergiant $\mu$ Cep is shown
  for comparison \citep{cas-4}. 
  Right: The 9.7\mic\ silicate feature profile in the novae V1370~Aql (black),
  V838~Her (red) \citep{smith} and V705~Cas (green) \citep{cas-2,cas-4}; tick
  marks are the wavelengths of AIR features in V705~Cas. The silicate feature in
  the red supergiant $\mu$ Cep (blue) is again shown for comparison.
  \label{cas4_sil}} 
  \end{figure}


Observations with {\it Spitzer} have further transformed our view of dust
formation and emission in novae. Fig.~\ref{v1065-cen} shows two contrasting
novae observed with {\it Spitzer:} V1065~Cen, which formed primarily silicate
dust \citep{helton-v1065}, and DZ~Cru, which produced amorphous carbon dust and
no (or negligible) silicate \citep{dzcru}. In both cases the presence of nebular
and fine-structure lines is evident but in the case of DZ~Cru, there are
additional broad features due to AIR emission (see below).

The broad dust continuum seen in Figs~\ref{cas93} and \ref{v1065-cen} is
generally attributed to carbon and the presence of AIR features is consistent
with this. It is likely that the dust is some form of hydrogenated amorphous
carbon (HAC), in which H atoms are attached to the surface of amorphous carbon
particles of size $\sim0.1\ldots1$\mic, or polycyclic aromatic hydrocarbons
(PAH) in which H atoms are attached to fused aromatic rings. However, as noted by 
\cite{ER-pah}, the lifetime of free-flying PAH molecules in the hard radiation
field of novae is extremely short and if such exist in the nova environment,
they may originate from the fragmentation of larger HAC particles. 

\subsubsection{The silicate features}

The 9.7\mic\ silicate feature arises from the stretching of the Si--O bond,
while the 18\mic\ feature arises from bending of the O--Si--O bonds. These
features do not therefore provide any diagnostic information (e.g. as to whether
the silicate is forsterite, Mg$_2$SiO$_4$, or fayalite, Fe$_2$SiO$_4$), which
requires observations at longer wavelengths. However the 9.7\mic\ feature in
V1370~Aql and V838~Her \citep[][respectively]{RDG-aql,smith} was broader than
was the case in V705~Cas and in evolved oxygen-rich stars such as $\mu$~Cep, and
peaks at longer wavelengths (see Fig.~\ref{cas4_sil}). This may be the result of
annealing of the silicate in the strong, hard radiation field of the nova.

Note the relative weakness of the 18\mic\ silicate feature in V705~Cas 320~days
after eruption (see Figs~\ref{cas93} and \ref{cas4_sil}). While this tends to be less prominent
than the 9.7\mic\ feature, it is normally seen in the IR spectra of evolved
oxygen-rich stars (cf. the 18\mic\ silicate feature in $\mu$~Cep in
Fig.~\ref{cas4_sil}). Laboratory  work by \cite{nuth} suggests that the 18\mic\
feature is relatively weak in freshly-condensed dust. Their experiments showed
that the ratio of the integrated absorption strength of the 9.7\mic\ feature to
that of the 18\mic\ feature decreases monotonically with increased processing by
annealing and oxidation; they suggested that this ratio could therefore be an
indicator of the age the silicate. Thus the relative weakness of the 18\mic\
feature in the young ejecta of V705~Cas is likely a pointer to the fact that the
dust is ``fresh'', and does indeed form in material ejected in the nova
explosion. That the ``9.7:18'' ratio is a potential indicator of silicate age is
consistent with the {\it Spitzer} IRS observations of V2362 Cyg, in which the
18\mic\ feature is initially (446 days after eruption) weak, but increases in
strength over the subsequent 400 days (see Fig.~\ref{v1065-cen}).


\subsubsection{The AIR features}\label{s:air}

AIR emission is frequently seen in novae (such as V842~Cen and V705~Cas; see
Fig.~\ref{cas93}, lower frames) displaying  a deep minimum in the visual light
curve; they were first seen in a nova in the case of V842~Cen \citep{HMcG}. In
most astrophysical sources (such as \pion{H}{ii} regions, planetary nebulae, AGB
stars etc.), AIR emission is attributed  to PAH molecules \citep[e.g.][]{tielens},
or HAC, although each might represent the extreme ends of a size spectrum of
carbonaceous dust grains.

  \begin{figure}
  \centerline{\includegraphics[angle=0,width=10cm]{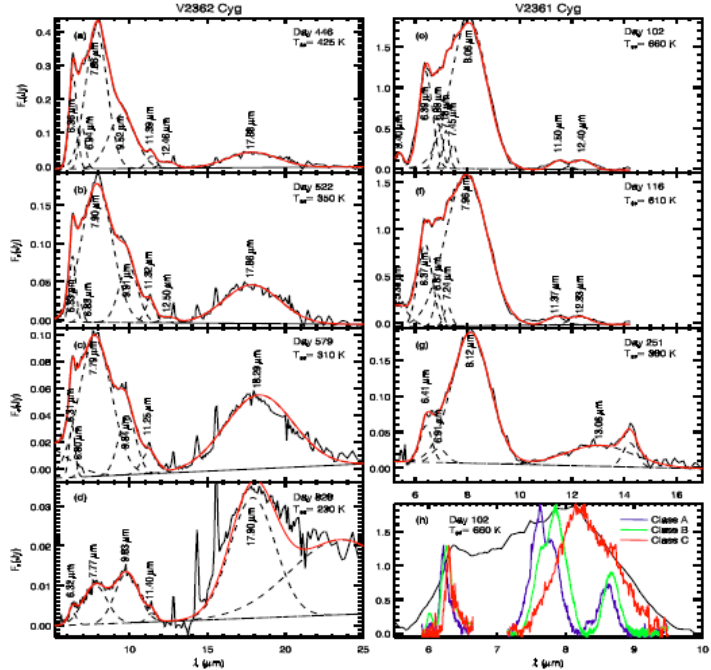}}
  \caption{{\it Spitzer} spectra of two dusty novae (V2362~Cyg, left, and
  V2361~Cyg, right) displaying AIR features (black curves); in each case a dust
  continuum has been subtracted. Dashed lines are individual AIR features,
  represented by gaussians; red curves are sums of individual components, fitted
  to the data. From \cite{helton-pah}. \label{kwok}} 
  \end{figure}  
  
Prominent AIR features are seen at 3.28\mic, 3.4\mic, 6.25\mic, 7.7\mic,
8.6\mic\ and 11.25\mic\footnote{There are other features at other, particularly
longer, wavelengths but we do not discuss these here.} and are much broader than
emission lines originating in the gas phase. The 3.28\mic\ feature is attributed
to C--H stretch in aromatic ($sp^2$) hydrocarbons, the 3.4\mic\ feature to
aliphatic ($sp^3$); the 7.7\mic\ feature is a blend of several C--C stretching
modes, while the 11.25\mic\ feature is due to C--H out-of-plane bending
\citep[see][for a review]{tielens}.

In novae in which AIR features have been observed, the 3.4\mic\ feature is much
stronger than the 3.28\mic\ feature (in contrast to the case in other
astrophysical sources), while the ``7.7'' and ``11.25'' features appear at
8.1\mic\ and 11.4\mic\ respectively. In view of the large over-abundance of
nitrogen in nova ejecta, the AIR carrier in novae
will almost certainly be ``contaminated'' by N, which will have a significant
effect on the properties of the AIR features \citep[see e.g.][]{cas-4}.

\cite{peeters} have classified AIR features according to their relative
strengths and novae AIR features seem to fit Peeters et al.'s ``Class~{C}''.
Objects in this class show no 7.7\mic\ feature but instead a feature at 8.2\mic;
moreover they also display a weak 6.25\mic\ feature, and an extremely weak
11.25\mic\ feature; these characteristics are very reminiscent of those seen in
novae. In evolved and Herbig Ae/Be
stars the wavelength of the ``7.7'' feature is highly environment-dependent,
with a clear dependence of wavelength on the effective temperature of the
exciting star \citep[e.g.][and references therein]{peeters,tielens}.

Objects in Peeters et al.'s Class~C are cool $(\sim5000$~K)
evolved stars, in stark contrast to the effective temperature of stellar remnant
of novae (expected to be $\sim100\,000$~K). It is curious that novae
display AIR features that are reminsicent of cool stars rather than of the hot
environment that one would expect. \cite{dzcru} have suggested that this
situation arises because the AIR carrier in novae has only recently been exposed
to the hard radiation field of the nova, having been hitherto been protected
within the dense clumps needed to form the dust in the first place, thus
limiting the photo-processing it has experienced. Alternatively, it may be that
only photons having lower energy ever penetrate deeply into the clumps, those
of higher energy being converted to lower energy by Lyman and similar
scattering at the clump surface, thus mimicking a low temperature environment.

The anomalous AIR features in novae are discussed by \cite{helton-pah}, who
present a detailed examination of the {\it Spitzer} spectra of V2361~Cyg and
V2362~Cyg (see Fig.~\ref{kwok}). In the case of V2362~Cyg the broad features
around 9.5\mic\ and 18\mic\ may be due to silicates, which are absent in
V2361~Cyg (see also above).

An alternative interpretation of the AIR features in novae is given by
\cite{kwok}, who note that the features appear in novae {\em as the dust is
formed}. They too have examined the {\it Spitzer} spectra of the two dusty novae
V2361~Cyg and V2362~Cyg and conclude that the AIR features in novae (and in
other sources) arise in complex organic solids which have a mixed $sp^2-sp^3$
composition. These materials (e.g. coals, kerogens) are present as an insoluble
residue in carbonaceous chondrite meteorites and display not only the usual AIR
features but  also very broad ``plateau'' features. 

\section{The recurrent novae} \label{s:recurrent}

A nova system that is observed to undergo a second eruption becomes a recurrent
nova: the difference between classical and recurrents is the selection effect
that a recurrent is a classical that has undergone more than one eruption
\citep[see][for recent reviews]{anupama,PASP-rsoph}. Few recurrents are known, and
those that are known form a very heterogeeous group. Other potential recurrents
have been identified, on the basis of the nature of their secondaries and the
amplitude at outburst \citep[see e.g.][and references therein]{weight,darnley}.

Since the widespread availability of IR in the 1970s there have been few RN
eruptions and so few IR observations of RNe in outburst exist, especially
longward of 3\mic. The first extensive IR observations were those of the 1985
eruption of RS~Oph \citep{rsoph-85}. In recent years, observations longward of
3\mic\ of the eruptions of RS~Oph (2006) and T~Pyx (2011) have been obtained and
we describe these here.

\begin{figure}
\centerline{\includegraphics[width=6cm]{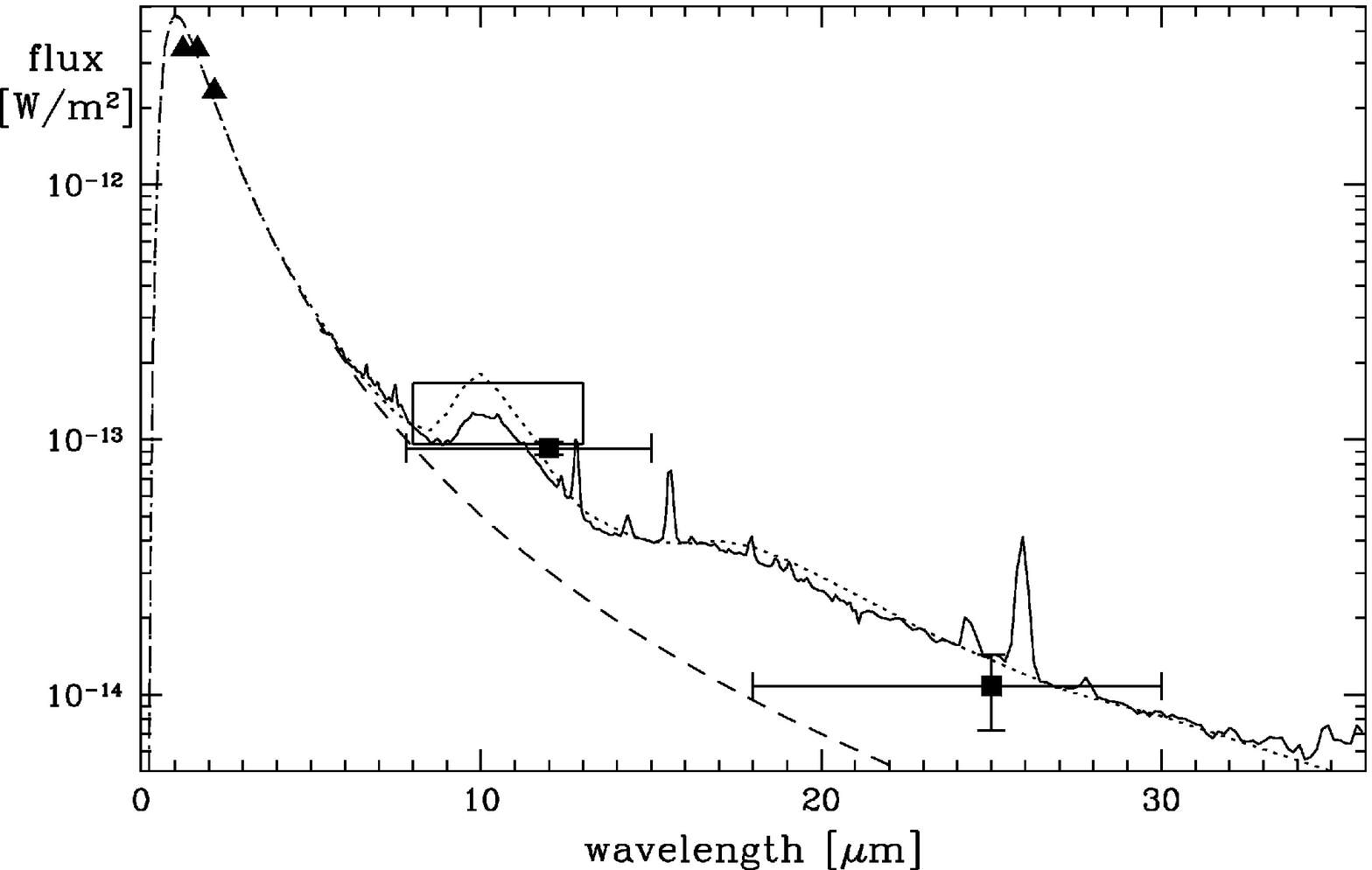} \qquad
            \includegraphics[width=6cm]{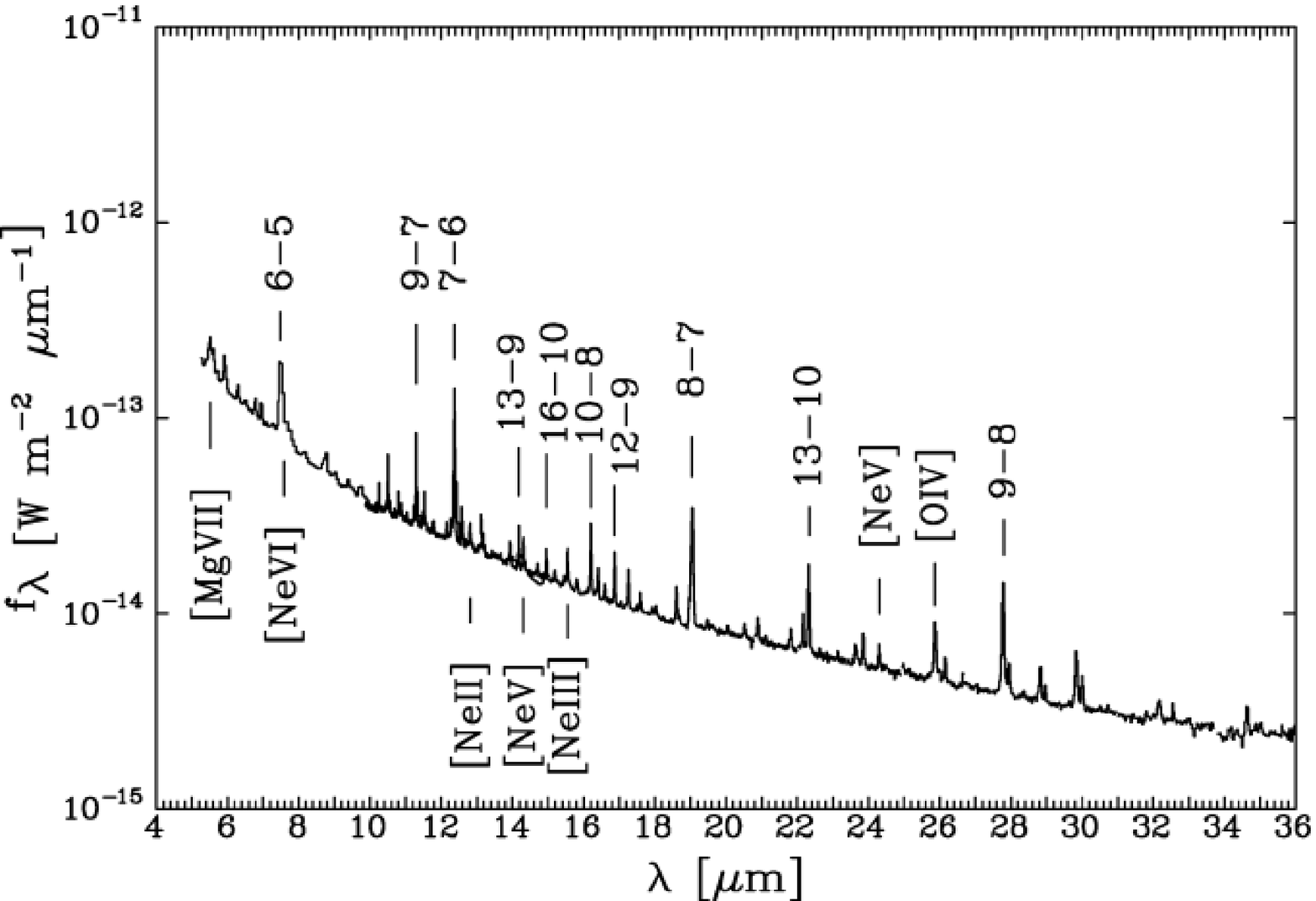}}
\medskip
\centerline{\includegraphics[width=6cm]{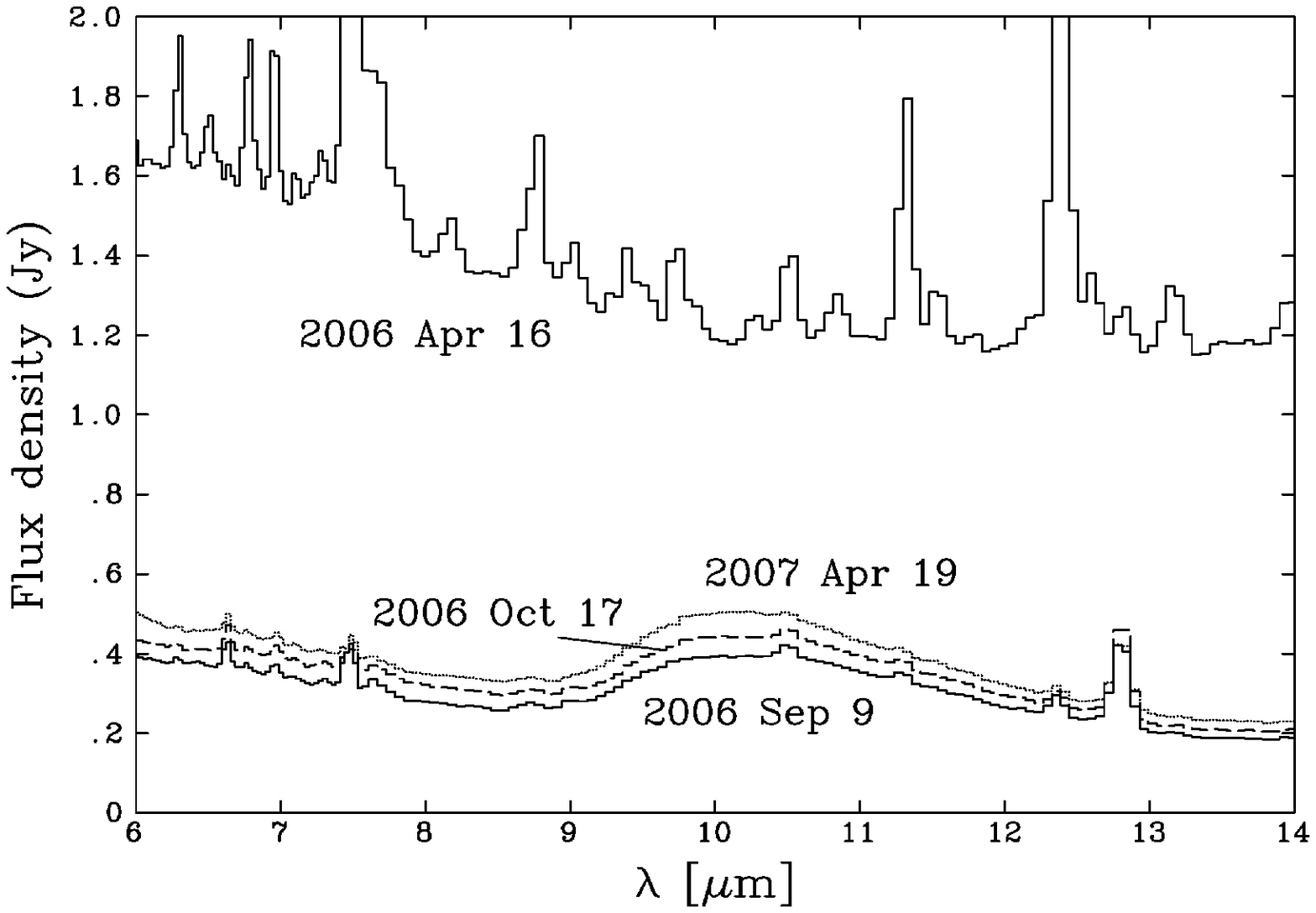} \qquad
            \includegraphics[origin=lt,width=6cm]{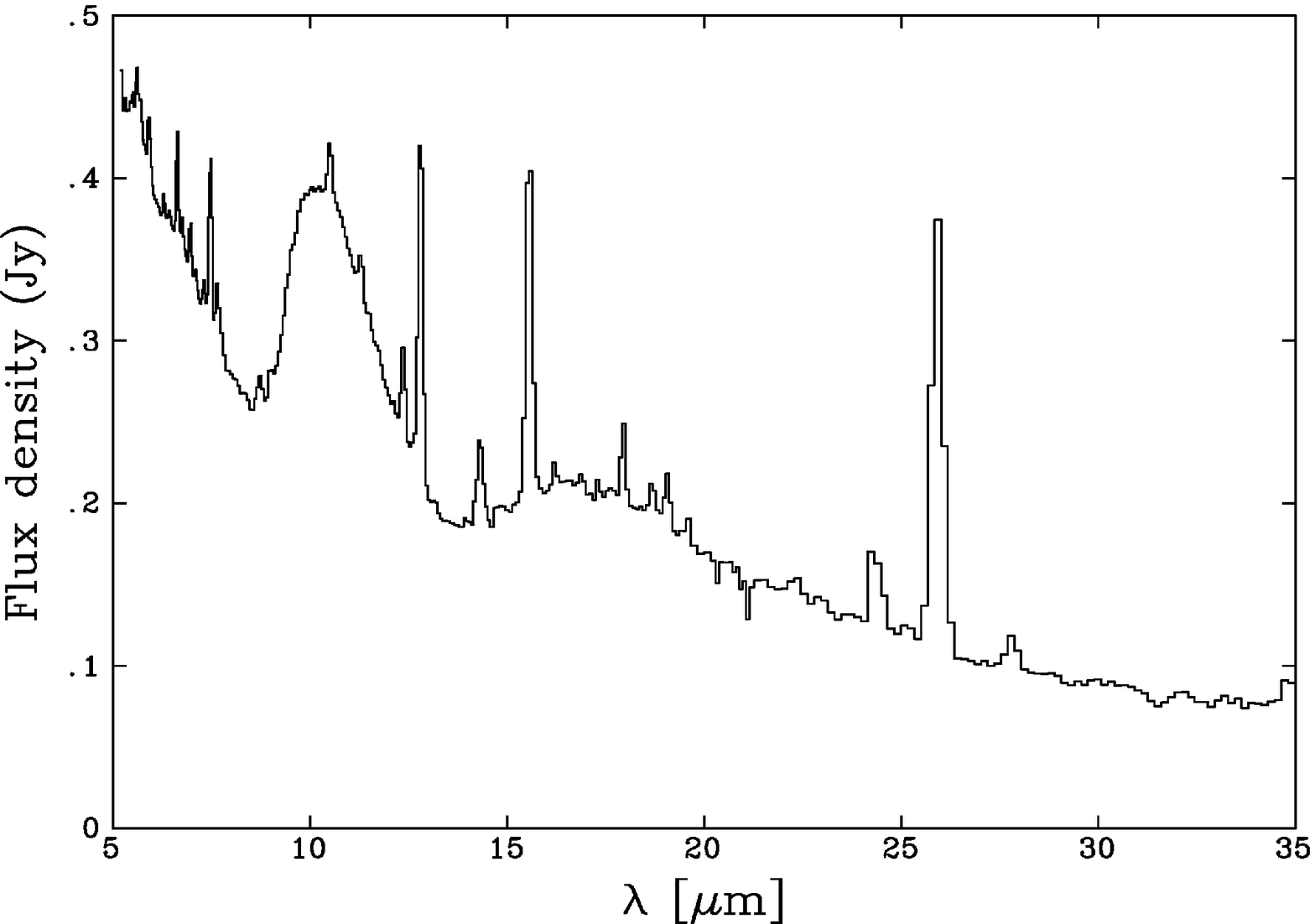}}
 \caption{Far-IR observations of RS~Oph. Top left: IRAS observations
 superimposed on {\it Spitzer} spectrum obtained following 2006 eruption
 \citep{vanloon}. Top right: emisison line spectrum during the 2006
 eruption, as observed by {\it Spitzer}. Botton frames: silicate dust
 in the environment of RS~Oph. {\it
 Spitzer} data from \cite{rsoph-1,rsoph-2}.\label{rsoph_spitzer}} 
\end{figure}

\subsection {RS~Oph (2006)}

{\it Spitzer} IRS observations of the 2006 eruption of RS~Oph are described by
\citep[][see Fig.~\ref{rsoph_spitzer}]{rsoph-1,rsoph-2}. An observation
62.5~days after outburst shows a rich emission line spectrum (including
fine-structure and coronal lines) super-imposed on a free-free continuum
\citep{rsoph-1}. The presence of coronal lines is not unexpected as the ejected
material from the 2006 eruption runs into and shocks the red giant wind
\citep{kahn}. However, unlike the case for classical novae, it is not straight-forward to
use the fine-structure and coronal lines to determine abundances as the observed
lines are a mix of shocked wind and ejecta. A first  requirement is the
determination of abundances in the wind, either by determining abundances in the
atmosphere of the red giant \citep[e.g.][]{pavlenko}, or in the red giant wind
\citep[e.g. by analysing the IR spectrum of the flash-ionised wind;][]{rsoph-0}.
Analysis of the coronal lines after the 2006 eruption indicates that there are two regions in the
emitting gas, with temperatures $1.5\times10^5$~K and $9\times10^5$~K.

As the free-free emission declined the {\it Spitzer} observations of RS~Oph
revealed a major surprise: a substantial amount of silicate dust in the
circum\-stellar environment \citep[see Fig.~\ref{rsoph_spitzer},
from][]{rsoph-2}. Note the strong 18\mic\ silicate feature, which indicates that
the dust observed cannot be freshly condensed as this would show a weak 18\mic\
feature \citep[see above;][]{nuth}. The relatively strong 18\mic\ feature
indicates that the dust is ``old'', and that there is a substantial amount of
dust in the environment of the RS~Oph binary that is completely oblivious to the
nova eruption.

RS~Oph was observed in the course of the IRAS survey in 1983, and {\it Spitzer} observations that show the silicate emission in the aftermath of the 2006 eruption are super-imposed on the IRAS data in Fig.~\ref{rsoph_spitzer}
\citep{vanloon}. That the IRAS data were obtained before either the 1985 or the 2006 eruptions reinforces the  fact that the dust is pre-existing, and a
constant feature of the RS~Oph system; a substantial fraction of the red giant
wind must \citep[as noted by][]{woodward}, be ``blissfully unaware'' of the
violent events that occur near it. This silicate dust presumably originates in
the wind of the red giant star in the RS~Oph binary system and its discovery has major implications for the evolution of the underlying binary, and for mass-loss from the system.
Given that the 2006 observations were the the first to see the
silicate, it is not known whether the 2006 erupion (and indeed previous
eruptions) affected the amount of dust (or indeed its nature) in the
circum\-stellar environment; observations of the next outburst in the
same wavelenth range will be valuable to see if the eruptions have any effect on the dust. 

\subsection{T~Pyx (2011)}

\begin{figure}
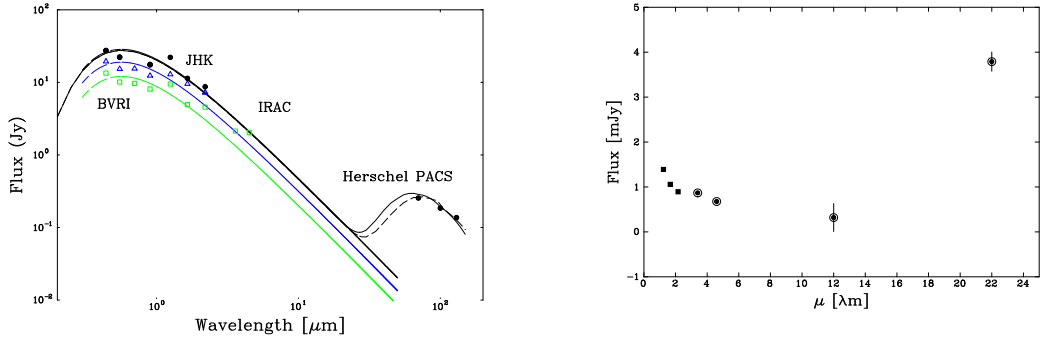

  \centerline{\includegraphics[angle=-90,width=7cm]{tpyx_SED.eps} \qquad
               \includegraphics[angle=-90,width=6.5cm]{tpyx_wise.eps}}
  \caption{Left: SED of T~Pyx following the 2011 eruption. Ground-based IR data
  from Mt Abu and SMARTS \citep{smarts}; mid-IR data from {\it Spitzer} and {\it 
  Herschel}. Spectral fits for 9000~K blackbody illuminating pre-existing circumstellar
  dust. From \cite{evans-tpyx}. Right: inter-outburst SED, based on 2MASS and WISE sky
  surveys. See text for details. \label{tpyx}} 
  \end{figure}

T~Pyx was the first nova to be identified as a recurrent, and has undergone
6~nova eruptions since 1890. Its most recent eruption (2011) was observed with
{\it Spitzer} and {\it Herschel}. The SED 26.2~days after the eruption is shown
in Fig.~\ref{tpyx} (black curve) and is consistent with emission by dust
longward  of $\sim50$\mic. \cite{evans-tpyx} conclude that the emission is due
to pre-existing dust in the environment of T~Pyx, the result of the sweeping up
of interstellar dust, either by winds or by material ejected in the course of
previous recurrent nova eruptions.
T~Pyx is the first nova in which an ``infrared echo'' -- postulated by \cite{BE2}
to account for the dust emssion -- has been observed. If this is indeed the case
the reverberation effect means that the far-IR emission should be detectable for
several years after the 2011 eruption \citep{evans-tpyx}.

T~Pyx is detected in the 2MASS \citep{2mass} and WISE \citep{wise} sky surveys and the SED based on these is
also shown in Fig.~\ref{tpyx}. There seems to be an excess at 22\mic, although observations
at other wavelengths would be valuable to confirm this. If so, the quiescent data seem
consistent with the presence of cool dust in the environment of T~Pyx between outbursts.

\section{Future prospects}

\subsection{The Stratospheric Observatory for Infrared Astronomy (SOFIA)}

\begin{figure}
  \centerline{\includegraphics[angle=0,width=8cm]{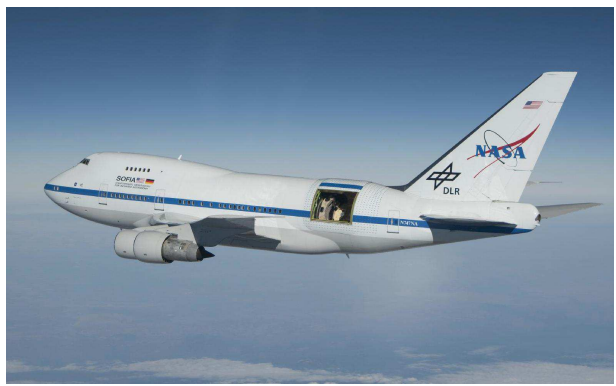}}
  \caption{NASA's SOFIA infrared observatory over Southern California's high
  desert on July 13, 2010 during flight testing in preparation for the Early
  Science Program. Image courtesy of the NASA Dryden Flight Research Center
  Photo Collection. \label{SOFIA}} 
  \end{figure}

The US/German Stratospheric Observatory for Infrared Astronomy (SOFIA), a joint
project of NASA and the Deutsches Zentrum f\"ur Luft und Raumfahrt (DLR), is a
gyro-stabilized 2.5-meter clear aperture IR telescope mounted in the aft
fuselage of a Boeing 747-SP aircraft \citep[see Fig.~\ref{SOFIA} and][]{young}. 
The observatory will make sensitive IR spectroscopic measurements at wavelengths
from 0.3\mic\ to 1.6~mm. SOFIA, which will fly until the
mid-2030s, will be a key facility for chemical/dynamical studies of classical
and recurrent novae. The characteristics of the initial suite of spectrometers
are summarized in Table~\ref{t:sofia}. 

SOFIA will fly at altitudes as high as 45,000 feet
(13.7~km), above 99.8\% of the remaining atmospheric water vapour. At this 
altitude, the precipitable atmospheric water
typically has a column depth of less than 10\mic, 20--100 times lower than at
good terrestrial sites. The atmospheric transmission averages 80\% or better
across SOFIA's wide wavelength range.

\begin{sidewaystable}
  \caption{SOFIA's First Generation Spectrometer Summary$^a$}\label{t:sofia} 
  \medskip
  \begin{center}
    \begin{tabular}{llllcc}\hline
    \multicolumn{1}{c}{Name$^b$} & \multicolumn{1}{c}{Description} & \multicolumn{1}{c}{PI} &
    \multicolumn{1}{c}{Institution} & Wavelengths &  Spectral \\ 
         &             &    &             &   (\mic)    &  Resolution  \\ \hline
   FORCAST & Mid-Infrared Camera and Grism Spectrometer & T. Herter &
   Cornell & 5--40 &  90--1200 \\
   GREAT & Heterodyne Spectrometer & R. G\"usten & MPIfR & 60--240 & 106--108 \\
   FLITECAM & Near-Infrared Camera and Grism Spectrometer &
   I. McLean &   UCLA &  1--5 & 900--1700 \\
   EXES & Mid-Infrared Spectrometer & M. Richter & UC Davis & 5--28 & 3000, 104,
   105 \\
   FIFI-LS &  Integral Field Far-Infrared Spectrometer & A. Krabbe &
   U Stuttgart & 42--210 & 1000--3750 \\
     \hline
    \end{tabular}\\[5pt]
    \begin{minipage}{15cm}
        \small (a) Details available at \underline
	{http://www.sofia.usra.edu/Sciecnce/instruments};\\
	(b)~Earliest availability for GI programmes:
	Cycle~1 (2013): FORCAST with grisms, and GREAT; Cycle~2 (2014): EXES and
	FIFI-LS. 
      \end{minipage}
  \end{center}
\end{sidewaystable}
  
SOFIA has several features that will make it an outstanding facility for observing nova explosions
\citep{RDG-sofia}:
\begin{enumerate}
\itemsep=0mm
\item Its mobility -- the ability to travel to any airfield that can handle a 747 on short notice
-- will facilitate the timely monitoring of the temporal development of nova events anywhere in the
sky. Coverage of the entire development of a single nova requires observational cadences that cover
events that develop on time-scales of days, weeks, and months. 
\item  The spectroscopic capabilities of the instruments summarised in Table~\ref{t:sofia} will
enable the recording of many forbidden lines obscured by the atmosphere from ground-based
observatories and unavailable to the spectrometers of other space missions. An assessment of the
strengths of these lines is necessary to determine accurate elemental abundances in nova ejecta.
\item High resolution spectroscopy of these
lines will provide a powerful probe of the physical conditions (e.g., density and temperature),
ionisation state, energetics, mass, and kinematics of the ejected gas through studies of the atomic
fine structure lines of \fion{O}{i} (63\mic\ and 145\mic), \fion{O}{iii} (52\mic, 88\mic),
\fion{O}{iv} (25.9\mic), \fion{C}{ii} (158\mic), \fion{S}{i} (26\mic), \fion{Si}{ii} (34\mic), and
\fion{S}{iii} (18.7\mic); see also Tables~\ref{FS} and \ref{t:coronal}. Furthermore, many forbidden lines of neon,
whose abundance is an indicator of progenitor mass and WD type, lie in obscured atmospheric bands that can be observed by SOFIA.
\item SOFIA is unique in that the spectral coverage and
resolution of its spectrometers is perfect for assaying the detailed mineralogy of nova dust, a task
that was only cursorily begun by Spitzer IRS observations. In particular it will be possible to make
a detailed study of silicate crystallinity \citep[see e.g.][]{fabian,molster} in novae that produce
silicate dust (see Section~\ref{s:dust}), and of AIR fetaures at wavelengths not reached by the {\it
Spitzer} IRS \citep[see][]{tielens}. Of particular interest are the far-IR AIR features expected to be
associated with the low-lying vibrational modes corresponding to the ``drum-head'' modes of large PAH
molecules. 
\end{enumerate}

SOFIA observing time for General Investigator (GI) programmes will be competed
for and awarded on annual cycles. Proposal awards for Cycle~1 
will be announced in late August, 2012 and observations will occur during 2012 December -- 2013
December. Cycle~2 proposals will be solicited in 2012 December -- 2013 Janurary, and will be
harvested in 2013 April for flight during 2014 January -- 2014 December. By 2014, the SOFIA
programme expects to be doing up to $\sim~1000$~hours of science observations per year (12\% of the
total time). Observing opportunities will be announced from time to time on the SOFIA website at
\underline{www.sofia.usra.edu}.

 \begin{figure}
  \centerline{\includegraphics[angle=0,width=6.5cm]{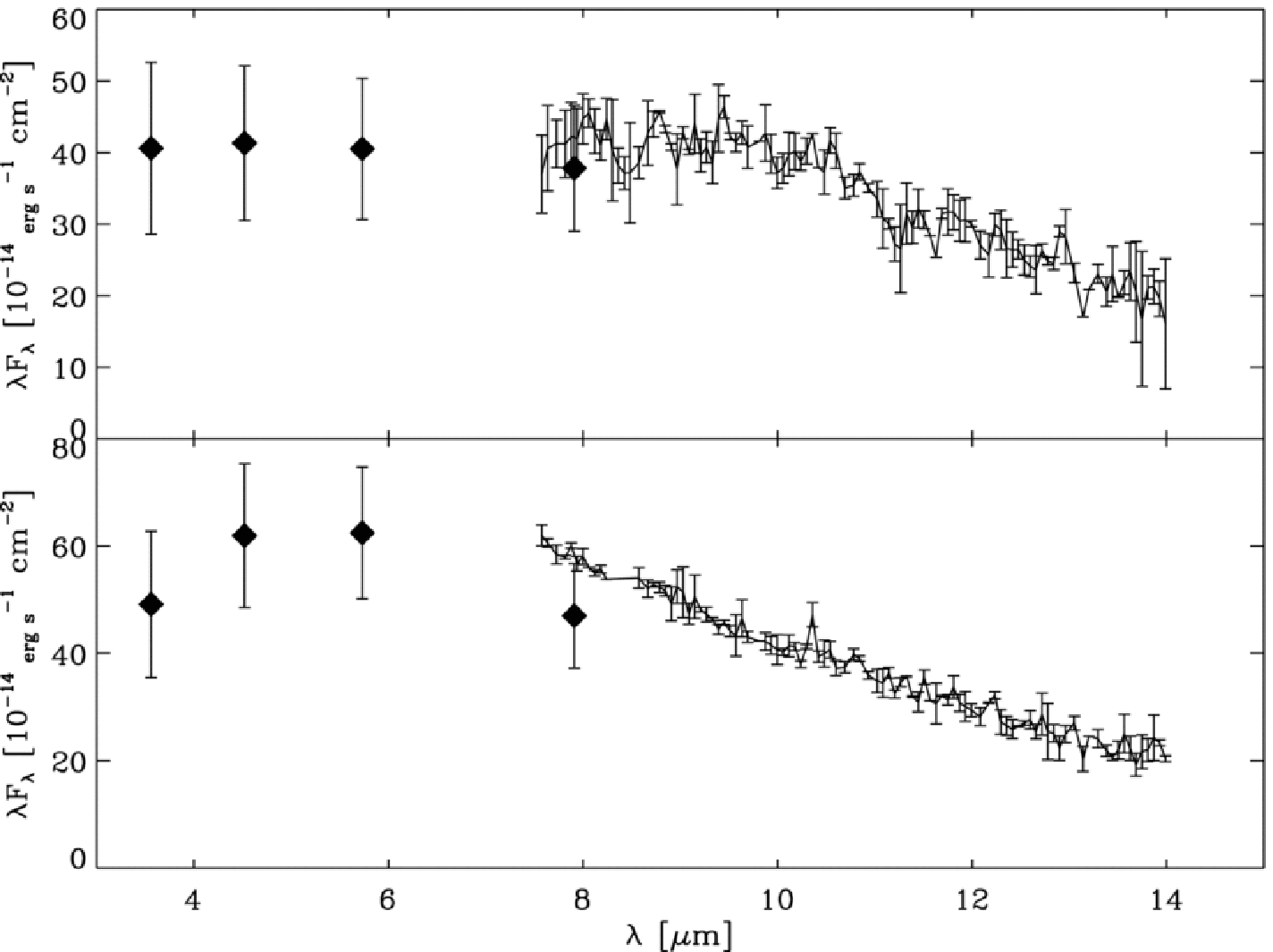} \qquad
               \includegraphics[angle=0,width=6.5cm]{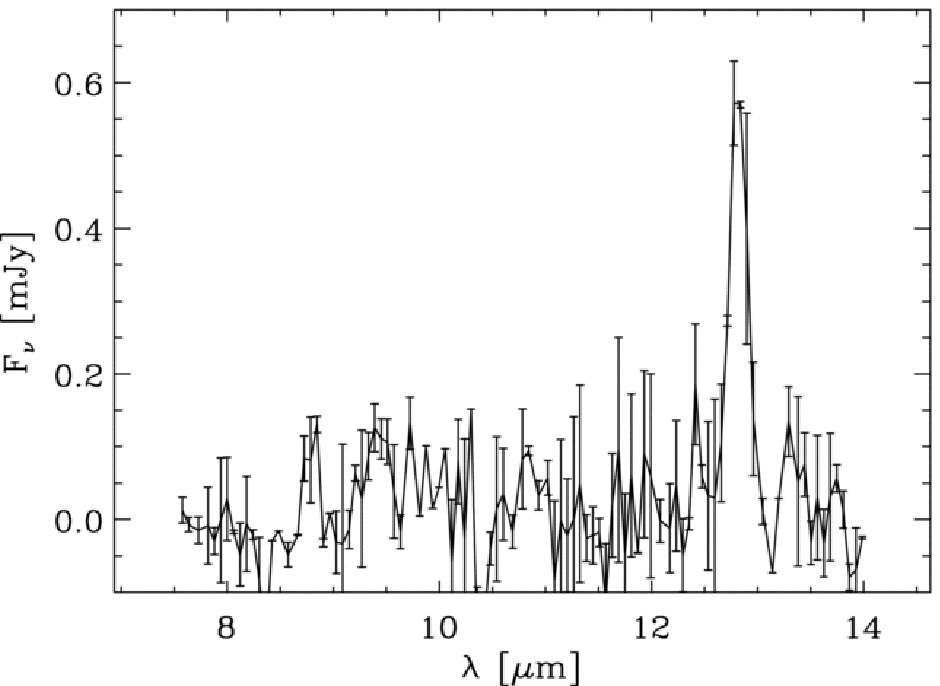}}
  \caption{Left: IR SEDs for two M31 novae, M31N~2006-09c (upper panel) and M31N~2006-10a (lower panel); note the IR excess, due to emission by dust, in M31N 2006-10a. Right: {\it Spitzer} IRS spctrum of M31N~2007-11e, showing the \fion{Ne}{ii}12.81\mic\ fine-structure line. From \cite{shafter}.
  \label{m31_spitzer}} 
  \end{figure}

\subsection{The James Webb Space Telescope}

NASA's James Webb Space Telescope \citep[JWST][]{jwst} is now working to a 2018 launch, with
an expected ten year lifetime. It will have two spectrometers capable of following the temporal
spectral development of classical and recurrent novae. The near-IR spectrometer (NIRSPEC) will
cover 1--5\mic\ with spectral resolutions of $R\sim100-3000$, and the mid-IR instrument (MIRI) will
cover 5--28\mic\ at $R\sim100-3000$. These instruments have integral field units and are not
especially efficient for viewing single point sources like Galactic novae. 

However, the very high sensitivity of these instruments make them well suited
for discovering nova populations in external galaxies. Their potential in this
regard is well illustrated by {\it Spitzer} InfraRed Array Camera
\citep[][]{irac} and IRS observations of novae in M31 by \cite{shafter}.
Fig.~\ref{m31_spitzer} shows the IR SEDs of the two novae M31N~2006-09c and
2006-10a. The IR excess, peaking in $\lambda{f}_\lambda$ at $\sim4$\mic, is
evident in the case of M31N 2006-10a; this nova had $t_2 \sim50$~days (in the
$R$ band), characteristic of dust-forming novae in the Galaxy. The nova
M31N~2007-11e showed clear evidence for emission in the
\fion{Ne}{ii}12.81\mic\ fine-structure line, the Humphreys$-\alpha$ 12.37\mic\ line being very weak or absent. While the presence of the \fion{Ne}{ii} line is
a necessary, but not sufficient, condition for classification as a neon nova,
there is clearly the potential to undertake unbiassed population studies of
extragalactic novae with JWST. This is all the more so as extinction -- even
less well known in extragalactic environments than it is in the Galaxy -- is a
relativley minor irritant in the IR.

On the other hand, viewing constraints imposed by the L-2 orbit will likely
prevent JWST from making timely target of opportunity observations of many
novae. The spectral coverage limits JWST nova observations to wavelengths
shorter than 28\mic.   

\subsection{Other Facilities}

A number of very large aperture telescopes capable of 3--25\mic\ ground-based follow-up of nova
events have recently come into operation or will soon begin full-time operations.  These include
8--10~m telescopes such as Gemini \citep{gemini}, VLT \citep{vlt}, Keck \citep{keck}, and
the Large Binocular Telescope \citep[LBT;][]{lbt}. In addition to their sensitive spectroscopic
capabilities, several of these telescopes are equipped with laser guide stars and adaptive optics
that can achieve spatial resolutions of $0.04-0.10''$ at wavelengths $>3$\mic\ that may enable
them to image young nova ejecta in some cases.

Future very large ground-based giant segmented mirror telescope (GSMTs) such as
the Giant Magellan Telescope \citep[GMT;][]{gmt}, the Thirty Meter Telescope
\citep[TMT;][]{tmt}, and the European Extremely Large Telescope
\citep[E-ELT;][]{eelt} will extend the spectroscopic and imaging capabilities
even further.

Prospective observers should consult
the websites of all of these facilities for information about progress and observing opportunities 

\section{Concluding remarks}

Over the past $\sim30$~years IR observations have proven to be pivotal in advancing our understanding of the
nova phenomenon, and pan-chromatic observations -- combining IR with optical, UV and X-ray observations --
has been especially effective (as in the case of RS~Oph for example). 
It is highly desirable that future novae (both classical and recurrent) continue to be observed over as
broad a wavelength range as possible, particularly as systematic observations of extragalactic novae will
soon be a real possibility.  

While great strides have been made in recent years, 
there remain some substantial gaps in our understanding (and IR observational coverage) of novae. For
example we have little IR spectroscopy of molecules in novae, with CO being the only molecule securely
detected in the IR to date \citep[see above,][and references therein]{ER-CN2}; molecular
rotational/vibrational can provide crucial information about isotope ratios, critical to TNR modelling. 

However we can expect that, in the next decade, SOFIA will fill in the IR spectroscopic gaps tantalisingly
opened up by {\it Spitzer}. In particular fine-structure lines beyond the {\it Spitzer} IRS range will
become accessible, particularly in old, low density, remnants (such as DQ~Her). Furthermore, it will be
possible for the first time to undertake a detailed mineralogical study of nova dust, particularly nova
silicates, and the way in which the silicate evolves as it is processed. 

To paraphrase Winston Churchill: {\it Spitzer} ``is not the end. It is not even the beginning of the end.
But it is, perhaps, the end of the beginning''. 

\section*{Acknowledgements}

RDG was supported by NASA and the United States Air Force.

\label{lastpage}
\end{document}